\documentclass[twocolumn]{aastex6}
\usepackage{color}

\usepackage{amsmath}
\usepackage{url}
\usepackage{hyperref}

\usepackage[normalem]{ulem}

\shorttitle{Where Do Water Masers Arise in the Andromeda Galaxy?}
\shortauthors{Amiri \& Darling}
\begin{document}
\title{Water Masers in the Andromeda Galaxy:  II.  Where Do Masers Arise?}

\author{Nikta Amiri\altaffilmark{1} \& Jeremy Darling}
\affil{Center for Astrophysics and Space Astronomy, Department of Astrophysical and Planetary Sciences, 
University of Colorado, Boulder, CO, 80309-0389, USA}
\altaffiliation[\altaffilmark{1}]{Current address:  Jet Propulsion Laboratory, M/S 238-600, 4800 Oak Grove Dr., Pasadena, CA 91109, USA; }
\email{nikta.amiri@jpl.nasa.gov}

\begin{abstract}
We present a comparative multi-wavelength analysis of water maser-emitting regions and non-maser-emitting luminous 24~$\mu$m star-forming regions in the Andromeda Galaxy (M31) to identify the sites most likely to produce luminous water masers useful for astrometry and proper motion studies.  Included in the analysis are Spitzer 24~$\mu$m photometry, Herschel 70 and 160~$\mu$m photometry, H$\alpha$ emission, dust temperature, and star formation rate.  
We find significant differences between the maser-emitting and non-maser-emitting regions:  water maser-emitting regions tend to be more IR-luminous and show higher star formation rates.  
The five water masers in M31 are consistent with being analogs of water masers in Galactic star-forming regions
and represent the high-luminosity tail of a larger (and as yet undetected) population.  
Most regions likely to produce water masers bright enough for proper motion measurements
using current facilities have already been surveyed, but 
we suggest three ways to detect additional water masers in M31:  
(1) Re-observe the most luminous mid- or far-IR sources with higher sensitivity than was used in the Green Bank
Telescope survey; 
(2) Observe early-stage star-forming regions selected by mm continuum that have not already been selected by their 24~$\mu$m emission, and 
(3) Re-observe the most luminous mid- or far-IR sources, and rely on maser variability for new detections.   
\end{abstract}


\section{Introduction}

Water masers can arise in star-forming regions, in shocks, in stellar atmospheres, and in the vicinity of massive black holes \citep[see reviews by][]{reid1981,elitzur1992,lo2005}.  They
can indicate specific physical conditions and provide high brightness temperature sources for precise astrometry and proper motion studies \citep[see review by][]{reid2014}.  
While water masers' presence and intensity cannot be predicted based on observed conditions
in any given physical setting (mostly due to nonlinear amplification of small-scale conditions and anisotropic emission), 
there is good observational evidence indicating where water masers are most likely to be observed.  In the Galaxy, for example, the 
water maser detection rate toward (ultra)compact \ion{H}{2} regions is typically 50\% or higher 
\citep[e.g.,][]{churchwell1990,urquhart2011}.

The utility of water masers for extragalactic proper motion studies has been demonstrated in the Local Group and in water maser disks 
associated with massive black holes \citep[e.g.,][]{brunthaler2005,humphreys2013}.  In the Local Group, the masers are associated with star formation and can be used to measure
systemic proper motions and proper rotation (also known as ``rotational parallax'').  This has been done for M33 and IC 10 
\citep{brunthaler2005, brunthaler2007}, but detected water masers were notably absent from the Andromeda Galaxy (M31) until recently 
\citep{sullivan1973,greenhill1995,imai2001,darling2011}.   The proper motion of M31 is a 
key quantity for Local Group dynamics \citep[e.g.,][]{loeb2005}, and while \citet{sohn2012} and \citet{vandermarel2012} 
obtained a constraint on the tangential velocity M31 of $\leq 34.3$~km~s$^{-1}$ ($1\sigma$)
using the {\it Hubble Space Telescope}, suggesting
a nearly radial Milky Way-Andromeda trajectory, a second completely independent and possibly more precise measurement is worthwhile
\citep[][]{darling2011,darling2016}.

Water masers in M31 have been difficult to find, in large part due to the low distance-dimmed flux density and 
due to the large areal size of the molecular disk:  the disk is too large in angular size and the masers 
are too faint to simply map the entire disk in a reasonable amount of observing time using current facilities.  A 
Green Bank Telescope (GBT) \footnote{The National Radio Astronomy Observatory is a facility of the National Science Foundation operated under cooperative agreement by Associated Universities, Inc.} survey of 506 22 $\mu$m-selected regions detected only five water masers \citep{darling2016}. 
The selection method is inefficient, and the survey is barely sensitive enough to detect the most luminous Galactic analog water masers
associated with star formation.  Given what we know about the star-forming regions in M31 in a pan-spectral sense, we can
(1) learn more about how and where luminous water masers arise, and (2) apply this knowledge to identify additional likely sites
of water maser emission in M31, improving detection statistics and making future surveys more efficient.  Water masers can show significant
peculiar motion and variability, so the detection of additional water masers would substantially improve proper motion and rotation measurements of M31 
and reduce systematic effects.  An enhanced astrometric network of water masers could enable the detection of the apparent expansion of
--- and thus the measurement of a geometric distance to --- M31 as it approaches the observer at $-$300 km s$^{-1}$ \citep{darling2011,darling2013}.  

In this paper, we present a comparative multi-wavelength analysis of 22 GHz water maser-emitting and non-maser-emitting 24~$\mu$m-luminous 
star-forming regions in M31.  We use {\it WISE}, {\it Spitzer}, and {\it Herschel}\footnote{Herschel is an ESA space observatory with science instruments provided by European-led Principal Investigator consortia and with important participation from NASA.} infrared continuum maps, maps of derived quantities such as 
star formation and dust temperature, and archival catalogs to examine the differences between maser-emitting and non-maser-emitting regions, 
to examine correlations between observable quantities among each population, 
and to constrain the parameter space most likely to produce detectable water masers.  
Section \ref{sec:survey} summarizes the GBT survey presented in detail in \citet{darling2016},  
Section \ref{sec:data} describes data sources and new measurements, 
Section \ref{sec:sample} refines the sample used in the analysis, 
Section \ref{sec:results} presents the results of the measurements and data collation, 
Section \ref{sec:analysis} examines trends and differences among the masers and non-masing regions, and 
Section \ref{sec:discussion} discusses the best approach to identifying new water masers in M31.  
Section \ref{sec:conclusions} highlights the main findings of this study.  

Throughout the manuscript, we assume a distance to M31 of 780 kpc when calculating luminosities from continuum or line flux measurements.

\section{The Green Bank Water Maser Survey of M31}\label{sec:survey}

\begin{figure*}
\center
\includegraphics[width=1.0\textwidth,trim=0 55 0 0,clip=true]{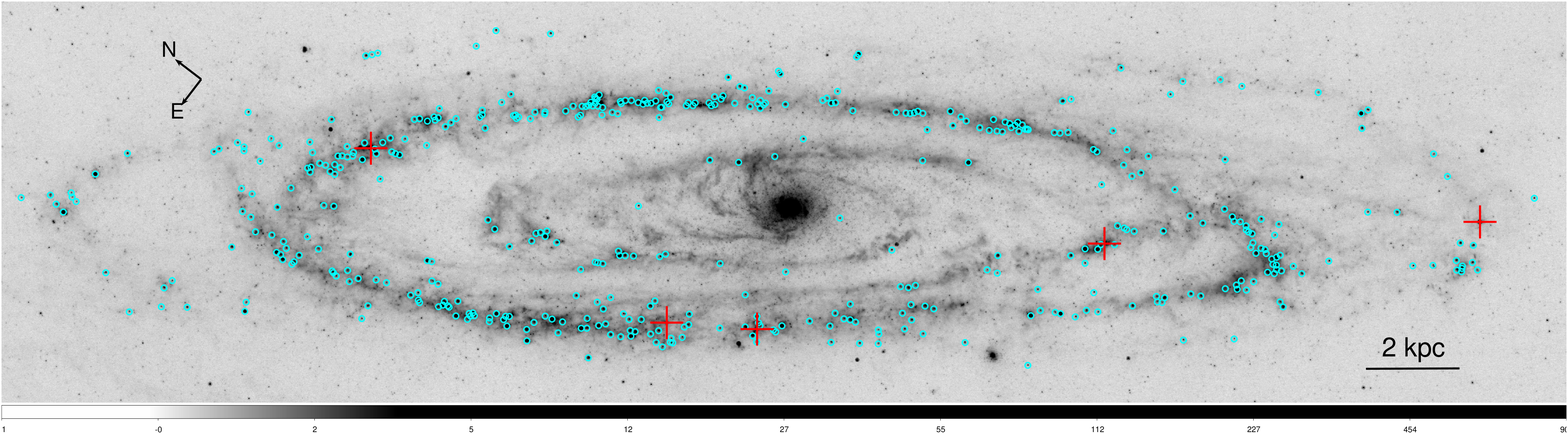}
\includegraphics[width=1.0\textwidth,trim=0 55 0 0,clip=true]{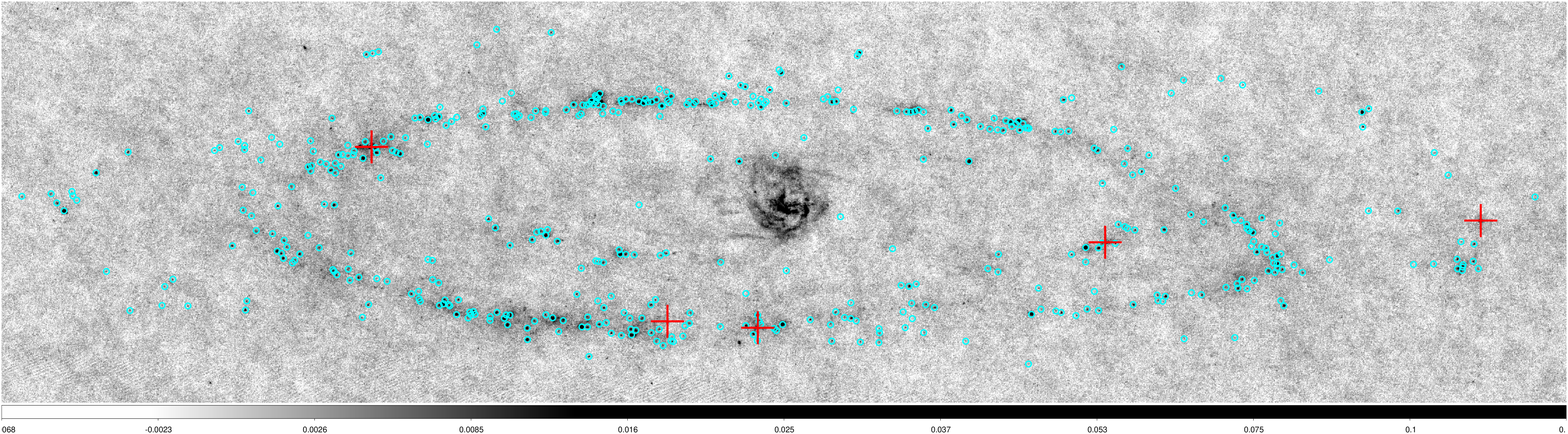}
\includegraphics[width=1.0\textwidth,trim=0 55 0 0,clip=true]{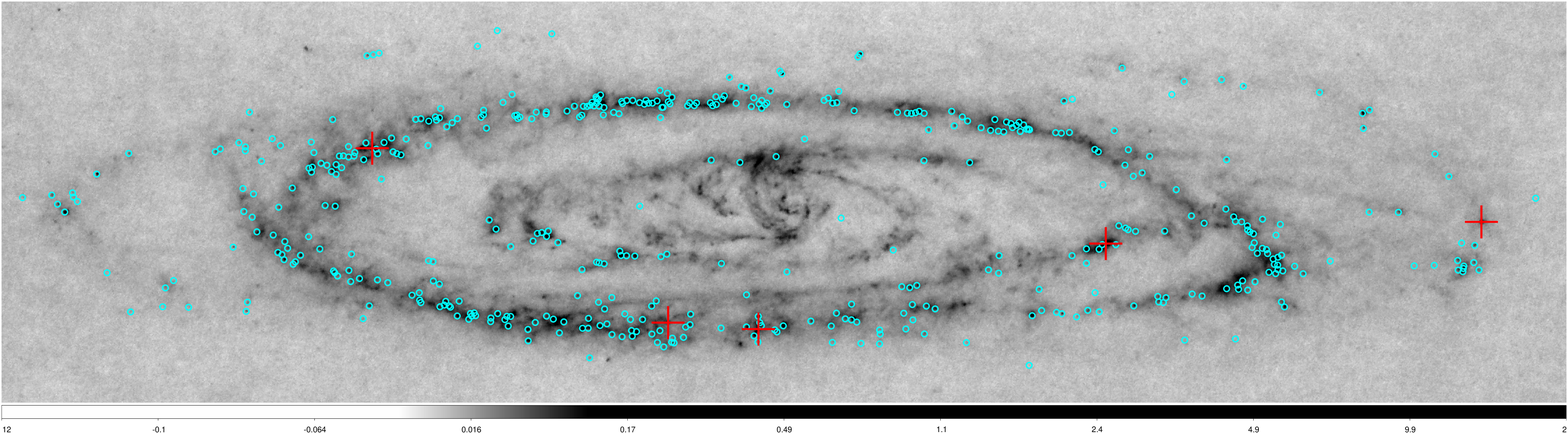}
\includegraphics[width=1.0\textwidth,trim=0 55 0 0,clip=true]{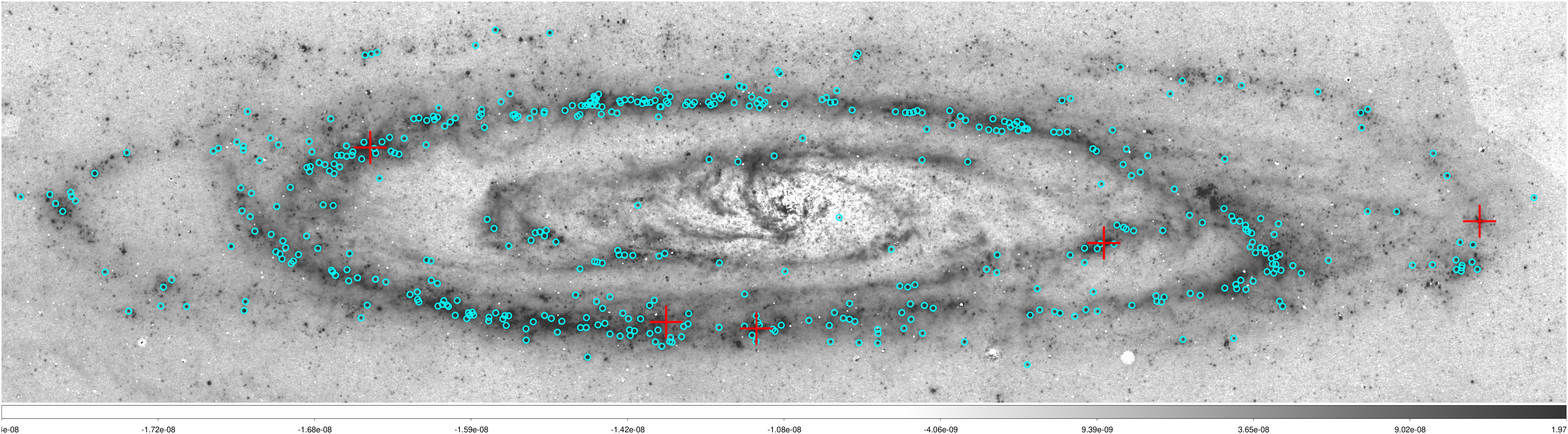}
\caption{ Infrared and star formation maps of M31.  Top to Bottom: Spitzer 24~$\mu$m \citep{gordon2006}, Herschel 70~$\mu$m,  Herschel 160~$\mu$m (Bruno Altieri, priv.\ comm.), and star formation rate \citep{ford2013}. Cyan circles show the 457 star-forming regions observed with the GBT \citep{darling2016}.  The circles are to scale, showing the 33\arcsec\ (125 pc) FWHM beam.  Red crosses indicate the location of the five detected water masers in M31 \citep{darling2011}.}
\label{fig:maps}
\end{figure*}

The water maser candidate selection for the Green Bank Telescope (GBT) survey for water masers in M31, 
the observing methods, data reduction, and results are presented in 
\citet{darling2011} and \citet{darling2016}.  In summary, 
we selected bright point sources from the Spitzer 24~$\mu$m map of M31 \citep{gordon2006}, and constructed a catalog of 506 objects from the brightest down to a point where most of the 24~$\mu$m emission becomes extended at about 4~MJy~sr$^{-1}$ (Figure \ref{fig:maps}, top). The compact 24~$\mu$m sources in M31 are likely associated with star-forming regions; strong water masers are known to arise in \ion{H}{2} regions in the Galaxy \citep[e.g.][]{walker1982}, and H$_2$O maser luminosity correlates with far-infrared (FIR) luminosity in Galactic 
star-forming regions as well as in star-forming galaxies \citep{felli1992,castangia2008}.

We observed the 6$_{16}-5_{23}$ 22.23508 GHz ortho-water maser line toward the 506 24~$\mu$m-selected 
regions in late 2010, late 2011, and early 2012 \citep{darling2011,darling2016}.  
Spectra were smoothed to 3.3 km~s$^{-1}$ channels, reaching an rms noise of $\sim$3 mJy in 
individual spectra and 0.17 mJy in a spectral mean stack of 299 objects aligned to the CO velocity \citep{nieten2006}.  Five water masers were 
detected \citep{darling2011}, and the detection rate after removing planetary nebulae and giant stars 
from the sample was 1.1(0.5)\% \citep[see Section \ref{subsec:interlopers} and ][]{darling2016}.  
The full details of the results of water maser observations, including the results of  NH$_3$ (1,1), NH$_3$ (2,2), and H66$\alpha$ observations,
are presented in \citet{darling2016}.  
In this paper, we use multi-wavelength data to investigate the physical and observed properties of water maser-emitting regions and to compare them to non-maser-emitting regions to understand where the water masers arise and how to detect additional water masers in M31.

\section{Multi-Wavelength Photometry and Derived Properties}\label{sec:data}

\floattable
\begin{deluxetable}{cccc}
\tabletypesize{\scriptsize}
\tablecaption{M31 Multi-Wavelength Data Sources \label{tab:data}}
\tablewidth{0pt}
\tablehead{
\colhead{Data} & \colhead{Resolution (\arcsec)} & \colhead{Telescope} & \colhead{Reference} 
}
\startdata
H$\alpha$ & 0.9--1.4&Mayall Telescope & \cite{azimlu2011}\\  
3.4 $\mu$m & 6.1 & {\it WISE} & \citet{wright2010}  \\
22 $\mu$m  & 22 & {\it WISE} & \citet{wright2010} \\
24 $\mu$m & 6 & {\it Spitzer} & \cite{gordon2006} \\
70 $\mu$m &  5.6 & {\it Herschel} &  \cite{groves2012}; B. Altieri (priv.\ comm.)\\
160 $\mu$m & 11.4 & {\it Herschel} & \cite{groves2012}; B. Altieri (priv.\ comm.) \\
\hline \rule{0pt}{3ex}
T$_{\rm dust}$ & 36 & {\it Herschel} and {\it Spitzer} &\cite{smith2012}\\
SFR & 6 & {\it Galex} and {\it Spitzer} &\cite{ford2013}\\
\enddata
\end{deluxetable}

\begin{figure*}
\center
\includegraphics[width=1.0\textwidth,trim=0 0 0 0,clip=true]{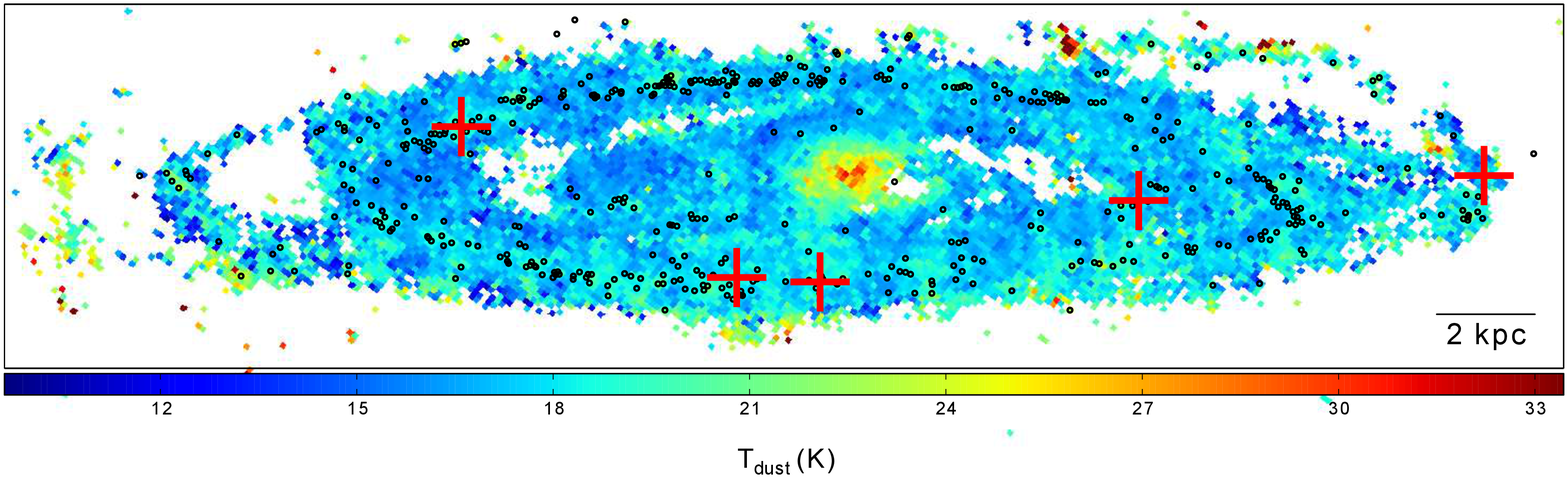}
\caption{Dust temperature map of M31 \citep{smith2012}. Black circles show the 457 star-forming regions observed with the GBT \citep{darling2016}. The circles are to scale, showing the 33\arcsec\ (125 pc) FWHM beam.  Red crosses indicate the location of the five detected water masers in M31 \citep{darling2011}.}
\label{fig:maps_dust}
\end{figure*}

\subsection{Data Sources}\label{subsec:data}
Table \ref{tab:data} summarizes the archival data used in the M31 water maser study, split into sources of photometry (H$\alpha$, mid- and far-IR) and 
derived quantities (dust temperature and star formation rate [SFR]). 
Figure \ref{fig:maps}  shows 24~$\mu$m, 70~$\mu$m, 160~$\mu$m, and star formation rate maps of M31, and 
Figure \ref{fig:maps_dust}  shows the dust temperature map \citep{smith2012}.  Both figures show the 
water masers and the non-detection locations.  

Spitzer observations of M31 at 24~$\mu$m were performed using the Multiband Imaging Photometer (MIPS) instrument with Point Spread Function (PSF) of 6\arcsec\ \citep{gordon2006}. The map covers an area of approximately 1$^{\circ}\times3^{\circ}$ oriented along the major axis of M31. The MIPS data analysis tool version 2.9 \citep{gordon2005} was used to produce the final mosaic map at 24~$\mu$m.

We obtained the Herschel maps of M31 at 70~$\mu$m and 160~$\mu$m from the public data of the Herschel archive \citep[][]{pilbratt2010,poglitsch2010}. The maps were re-processed by Bruno Altieri (ESA; private communication) with unimap map-maker \citep{piazzo2013}. The observations were performed using the Photodetector Array Camera and Spectrometer (PACS) instrument. Full details of the observing strategy can be found in \citet{groves2012}. The maps cover an area of roughly 1$^{\circ}\times 3^{\circ}$. The FWHM angular resolution of the 70~$\mu$m and 160~$\mu$m maps is 5.6\arcsec\ and 11.4\arcsec, respectively. 

\cite{smith2012} constructed the dust temperature map of M31 via pixel-by-pixel analysis of Spitzer and Herschel maps in the wavelength range 70--500 $\mu$m. All of the maps were convolved to the resolution of Herschel 500 $\mu$m map that has the largest FWHM resolution (36\arcsec).  The dust temperature for each pixel was measured by fitting a FIR through submillimeter spectral energy distribution with a single-temperature modified blackbody model:  $S_\nu = \kappa_\nu\, M_{\rm dust}\, B(\nu, T_{\rm dust})/D^2$, where $\kappa_{\nu}$ is the dust absorption coefficient described by a power law with dust emissivity index $\beta$ such that $\kappa_{\nu} \propto \nu^{\beta}$, $M_{\rm dust}$ is the dust mass with dust temperature $T_{\rm dust}$, $B(\nu, T_{\rm dust})$ is the Planck function, and $D$ is the distance to the galaxy. The estimated uncertainty in the dust temperature is 1.4 K. The dust temperature was measured where the fluxes in all six bands (five Herschel and MIPS 70~$\mu$m) had a signal-to-noise
ratio greater than 3$\sigma$.

The total star formation rate map of M31 (dust-obscured and unobscured) was constructed from the GALEX FUV and Spitzer 24~$\mu$m maps by \citet{ford2013}.  The contribution from the giant stellar population at 24~$\mu$m was removed using the IRAC Spitzer 3.6~$\mu$m band (see also Section \ref{subsec:interlopers}). 

We also used the optically identified \ion{H}{2} region catalog of \citet{azimlu2011} for this study.  
\citet{azimlu2011} used the data from the Nearby Galaxies Survey of \citet{massey2006}, which includes H$\alpha$ and R-band mosaics of ten overlapping fields across the disk of M31. \citet{azimlu2011} identified 3961 \ion{H}{2} regions above a 10$\sigma$ H$\alpha$ flux limit of 10$^{-16}$~erg~cm$^{-2}$~s$^{-1}$.
 
Finally, we obtained Wide-field Infrared Survey Explorer (WISE) maps of M31 at 3.4~$\mu$m (Figure \ref{fig: wise_noHII}, top) and 22 $\mu$m from the NASA/IPAC Infrared Science Archive\footnote{\url{http://hachi.ipac.caltech.edu:8080/montage}}. WISE mapped the sky in four bands at  3.4, 4.6, 12, and 22  $\mu$m with an angular resolution of 6.1\arcsec, 6.4\arcsec, 6.5\arcsec, and 12.0\arcsec, respectively \citep[][]{wright2010}.

\subsection{Photometry}\label{sec:photometry}
Photometric measurements at 24~$\mu$m, 70~$\mu$m, and 160~$\mu$m were performed using the Aperture Photometry Tool \citep[APT, ][]{laher2012}. The point spread function (PSF) FWHM of the 24~$\mu$m, 70~$\mu$m, and 160~$\mu$m images is 6\arcsec, 5.6\arcsec, and 11.4\arcsec, respectively, and the pixel size is 1.24\arcsec, 3.2\arcsec, and 6.4\arcsec.  We select aperture radii of 6.2\arcsec, 6.4\arcsec, and 6.4\arcsec. We chose similar aperture sizes at all wavelengths in order to match physical sizes in the photometry.

We performed aperture photometry on the SFR map using an aperture radius of 6\arcsec. The dust temperatures were obtained from the dust temperature map at each 24 $\mu$m source position (Figure \ref{fig:maps}). Five regions 
did not meet the 3$\sigma$ dust temperature threshold (Section \ref{subsec:data}) and were therefore omitted from the analysis sample (Section \ref{sec:sample}).

\floattable
\begin{deluxetable}{llllrrlrrr}
\tabletypesize{\scriptsize}
\tablecaption{Multi-Wavelength Properties of Water Maser Hosts in M31 \label{tab:maser}}
\tablewidth{0pt}
\tablehead{
\colhead{Maser} & \colhead{T$_{\rm dust}$} & \colhead{log(SFR)} & \colhead{log(24 $\mu$m)} & \colhead{log(70 $\mu$m)} & \colhead{log(160 $\mu$m)} & \colhead{log(H$\alpha$)} & \colhead{H$_2$O } & \colhead{L$_{\rm H_{2}O}$} & \colhead{log(L$_{\rm TIR}$)} \\
\colhead{(J2000)}  &\colhead{(K)} & \colhead{($M_{\odot}$ yr$^{-1}$)} & \colhead{(Jy)} & \colhead{(Jy)} & \colhead{(Jy)} & \colhead{(mW m$^{-2}$)} & \colhead{(mJy km s$^{-1}$)}  & \colhead{(L$_{\odot}$)} & \colhead{(L$_{\odot}$)}   
}
\startdata
003918.9+402158.4  &   23.0   & $   -4.1203  (     7   ) $ & $    -1.2372   (        7  ) $ & $     -0.136 (  10   ) $ & $    0.096    (      7  ) $ & $       -12.485     (       1    )  $ & $   447( 43)$   & 0.0063(6)   &    6.102(5)    \\
004121.7+404947.7  &   18.3   & $   -4.3142  (    11   ) $ & $    -1.3780   (       10  ) $ & $     -0.273 (  14   ) $ & $    0.270    (      5  ) $ & $       -13.445     (      66    )  $ & $ 95 ( 20 ) $   &    0.0013(3)   &    6.139(5)	 \\
004343.9+411137.6  &   19.4   & $   -4.3600  (    12   ) $ & $    -1.4511   (       12  ) $ & $     -0.413 (  19   ) $ & $    0.124    (      7  ) $ & $       -13.496     (      30    )  $ & $ 58 ( 9  ) $    &    0.00081(13) &      6.003(6)	 \\
004409.5+411856.6  &   18.4   & $   -4.8241  (    35   ) $ & $    -1.8660   (       32  ) $ & $     -0.466 (  21   ) $ & $    -0.097   (      11 ) $ & $      <  -16   	     $ &  $ 82 ( 14) $  &    0.0011(2)  &     5.811(10)  \\
004430.5+415154.8  &   22.6   & $   -3.9200  (     4   ) $ & $    -0.9897   (        4  ) $ & $     0.097  (  6    ) $ & $   0.361     (     4   ) $ & $      -12.370      (      2     )  $ & $ 67 (15)$    &    0.0009(2)  &     6.353(3)\\
\enddata
\tablecomments{The integrated H$_2$O maser flux densities and luminosities were obtained from \citet{darling2011}.
Parenthetical values indicate 1$\sigma$ statistical uncertainties. The 1$\sigma$ uncertainties for photometric flux densities and SFR indicate statistical uncertainties for images with high signal to noise ratios, but the systematic uncertainties are likely to be higher.}
\end{deluxetable}

Due to crowding in the molecular ring, estimation of the local background is difficult. We subtract a local non-annulus sky background using the default ``Model F'' algorithm in the APT that estimates the sky background using bilinear interpolation of the mode statistic. This model has been suggested for photometry in crowded fields \citep{laher2012}. 

Although re-scaling all maps to the largest resolution of 11.4\arcsec\ at 160 $\mu$m would be appropriate to obtain photometry over a uniform physical scale, we chose to perform photometry at the original resolution of the maps. This is due to the fact that the resolution of 24 $\mu$m, 70 $\mu$m, and SFR maps are similar and in the range 5.6\arcsec--6\arcsec. Since the objects are in a crowded field, re-scaling the maps to a larger resolution would lead to (additional) confusion.

We obtained the encircled energy fraction (EEF) for the Herschel images from the PACS Photometer Point-Source Flux Calibration 
document\footnote{\url{http://herschel.esac.esa.int/twiki/pub/Public/} \url{PacsCalibrationWeb/pacs_bolo_fluxcal_report_v1.pdf}}; the estimated aperture correction factor (1/EEF) for aperture radii of $\sim$6.4\arcsec\ (70 $\mu$m) and $\sim$6.4\arcsec\ (160 $\mu$m) corresponds to $\sim$1.56 and 2.6, respectively. For the 24 $\mu$m map, we adopt an aperture correction factor of $\sim$1.61 for the 6\arcsec\ aperture radius\footnote{\url {http://irsa.ipac.caltech.edu/data/SPITZER/docs/} \url{mips/mipsinstrumenthandbook/50/}}.

The uncertainties assigned to the measured photometric flux densities correspond to the standard deviation of the photometric flux of a large number of blank sources in each image. We obtain aperture photometric flux for 50 blank sky locations and measure the standard deviation of the photometric flux of the blank sources; this gives a good measure of the true photometric error of the targets. The estimated 1$\sigma$ uncertainties in the 24~$\mu$m, 70~$\mu$m, and 160~$\mu$m maps correspond to 1.16$\times$10$^{-4}$, 0.0168, 0.0186 Jy, respectively. The 1$\sigma$ uncertainty for star formation rate corresponds to 1.2$\times 10^{-7}$ $M_{\odot}$ yr$^{-1}$. The 1$\sigma$ uncertainties for photometric flux densities and SFR represent statistical uncertainties for images with high signal to noise ratios, and the systematic uncertainties are likely to be higher. Measured flux densities and SFR for the water maser and non-maser sample regions are shown in Table \ref{tab:maser} and Table \ref{tab: photometry}, respectively.

The multi-wavelength data used in this work were obtained at different resolutions. While the resolution of the Spitzer and Herschel maps ranges from 5.6\arcsec\ to 11.4\arcsec, the resolution of the dust temperature map is $\sim$36\arcsec\ based on the resolution of Herschel maps at longer wavelengths \citep[e.g., 500 $\mu$m,][]{smith2012}. Additionally, the crowded field and the large PSF of Spitzer and Herschel maps (5.6\arcsec--11.4\arcsec) may introduce contamination from nearby or confused sources \citep[e.g.,][]{calzetti2005}.

\begin{deluxetable*}{lcllccll}
\tabletypesize{\scriptsize}
\tablecaption{Multi-Wavelength Properties of the Non-Maser Sample \label{tab: photometry}}
\tablewidth{0pt}
\tablehead{
\colhead{Object} & \colhead{T$_{\rm dust}$} & \colhead{log(SFR)} & \colhead{log(24 $\mu$m)} & \colhead{log(70 $\mu$m)} & \colhead{log(160 $\mu$m)} & \colhead{log(H$\alpha$)} & \colhead{log(L$_{TIR}$)}\\
\colhead{(J2000)}  &\colhead{(K)} & \colhead{($M_{\odot}$ yr$^{-1}$)} & \colhead{(Jy)} & \colhead{(Jy)} & \colhead{(Jy)} & \colhead{(mW m$^{-2}$) } & \colhead{(L$_{\odot}$)}}
\startdata 																					003852.5+401904.9 &    \nodata    & $    -5.0137  (    54   ) $ & $   -2.0871  (     53  ) $ & $     < -1.3              $ & $   <-1.2                $ & $      < -16           $ & 	  \nodata          \\
003904.8+402927.4 &   15.8    & $    -5.1991  (    82   ) $ & $   -2.2906  (     85  ) $ & $     -1.276      (  137  ) $ & $   -0.757    (    50  )   $ & $      -14.149     (    21   ) $ &$	 5.133   (  45	 )$ \\
003906.7+403704.5 &   21.7    & $    -4.8866  (    40   ) $ & $   -2.0037  (     44  ) $ & $     -0.914      (   59  ) $ & $   -0.458    (    25  )   $ & $      -14.613     (    18   ) $ &$   5.446    ( 21	 )$ \\
003909.8+402705.0 &   17.4    & $    -5.0078  (    53   ) $ & $   -2.3951  (    108  ) $ & $     < -1.3              $ & $   -0.724    (    46  )   $ & $      -13.146     (     3   ) $ &	  \nodata  	   \\
003910.2+403725.6 &   21.3    & $    -4.1747  (     8   ) $ & $   -1.2574  (      8  ) $ & $     -0.189      (   11  ) $ & $    0.116    (     7  )   $ & $     -13.157      (    1    ) $ &$	 6.091   (   4	 )$ \\
003914.6+404157.9 &   18.3    & $    -5.2161  (    86   ) $ & $   -2.3364  (     94  ) $ & $     < -1.3              $ & $   -0.749    (    49  )   $ & $      -13.668     (     5   ) $ &	  \nodata    	   \\
003916.1+403629.5 &   19.6    & $    -4.5794  (    20   ) $ & $   -1.6420  (     19  ) $ & $     -0.536      (   25  ) $ & $   -0.331    (    19  )   $ & $      < -16               $ &$	 5.690   (  12	 )$ \\
003930.2+402106.4 &   18.7    & $    -4.6986  (    26   ) $ & $   -1.8281  (     29  ) $ & $     -0.634      (   31  ) $ & $   -0.302    (    17  )   $ & $      -13.788     (     9   ) $ &$	 5.642   (  13	 )$ \\
\enddata
\tablecomments{Table \ref{tab: photometry} is published in its entirety in the 
electronic edition of the {\it Astrophysical Journal}.  A portion is 
shown here for guidance regarding its form and content. Parenthetical values indicate 1$\sigma$ statistical uncertainties. The uncertainty in T$_{\rm dust}$ is 1.4 K \citep[][]{smith2012}. The 1$\sigma$ uncertainties for photometric flux densities and SFR indicate statistical uncertainties for images with high signal to noise ratios, but the systematic uncertainties are likely to be higher. }
\end{deluxetable*}

\subsection{Optical Counterparts}\label{sec:optical}

We cross-matched the GBT survey sample with the \citet{azimlu2011} catalog of H$\alpha$ flux-limited optically 
identified \ion{H}{2} regions.  
We identified 346 H$\alpha$ counterparts in the non-maser catalog using a positional uncertainty of 10\arcsec.
We also found four (out of five) H$\alpha$ counterparts in the water maser sample.
The H$\alpha$ fluxes for the water maser and non-maser sources are listed in Table \ref{tab:maser} and Table \ref{tab: photometry}, respectively.

\section{The Study Sample}\label{sec:sample}

Not all objects in the GBT survey are \ion{H}{2} regions, and not all \ion{H}{2} regions in the survey are detected or measured 
in all properties used in the comparative analysis of maser- and non-maser-emitting regions.  Here we present the process
used to exclude planetary nebulae and giant stars from the sample \citep[also excluded from the detection statistics presented in][]{darling2016},
and we present the reduced study sample that has measurements of all quantities presented in Tables \ref{tab:maser} and \ref{tab: photometry}.

\subsection{Planetary Nebulae and Stellar Populations}\label{subsec:interlopers}
Although star forming regions emit strong 24~$\mu$m emission, they are not the only luminous 24~$\mu$m sources in M31. There are sources which emit significant 24 $\mu$m emission and are not associated with star forming regions, and these sources must be removed from the original sample of 506 sources. 

Planetary Nebulae (PNe) can emit strong infrared radiation at 24~$\mu$m and may represent a small fraction of the source sample. \citet{merrett2006} present the results of a survey of 3300 emission line sources in M31 observed with the Planetary Nebulae Spectrograph. After removing the extended emission from HII regions and background galaxies, they identify 2615 PNe candidates in M31. We cross matched our source list with the catalog of PNe and identified nine PNe candidates in our sample and removed them from the rest of the analysis (Table \ref{tab:removed}).

The giant stellar populations in M31 can also produce significant 24~$\mu$m emission. Red supergiants and asymptotic giant branch (AGB) stars show strong 3.4 $\mu$m emission that originates in thick circumstellar shells \citep[e.g.][]{barmby2006,mould2008}. In a recent study of dust heating in M31 using Herschel data in the wavelength range 70--500~$\mu$m, \citet{groves2012} find that ``old'' stellar populations (of Gyr age) can emit significant infrared radiation. \cite{ford2013} determine the effect of these stars on the apparent star formation rate from 24~$\mu$m map by measuring the ratio of $\alpha = I_{24}/I_{3.4}$ in regions where there is no active star formation, where $I_{24}$ and $I_{3.6}$ indicate the 24~$\mu$m and 3.6 $\mu$m intensity. They found significant correlation between 24~$\mu$m and 3.6~$\mu$m emission in the center of M31 predominantly from giant stars, and use $\alpha = 0.1$ to remove this component of the total 24~$\mu$m emission.

We examine the association of our sample with stellar populations by comparing the flux density of the sample at 3.4~$\mu$m to that at 22~$\mu$m. Aperture photometry was performed for the sources in the sample on the 3.4~$\mu$m and 22 $\mu$m WISE maps. We utilized the prescription for photometry described for WISE images in the user's guide to the WISE Preliminary Data Release \footnote{\url{http://wise2.ipac.caltech.edu/docs/release/prelim/} \url{expsup/sec2_3f.html}}. We obtained uncertainties in flux densities by performing aperture photometry on 50 blank regions in each map and measuring the standard deviations in aperture photometric flux. The 3$\sigma$ uncertainties at 3.4 $\mu$m and 22 $\mu$m are 0.16 and 0.18 mJy, respectively.

\begin{figure*}
\epsscale{1}
\center
\includegraphics[width=0.9\textwidth]{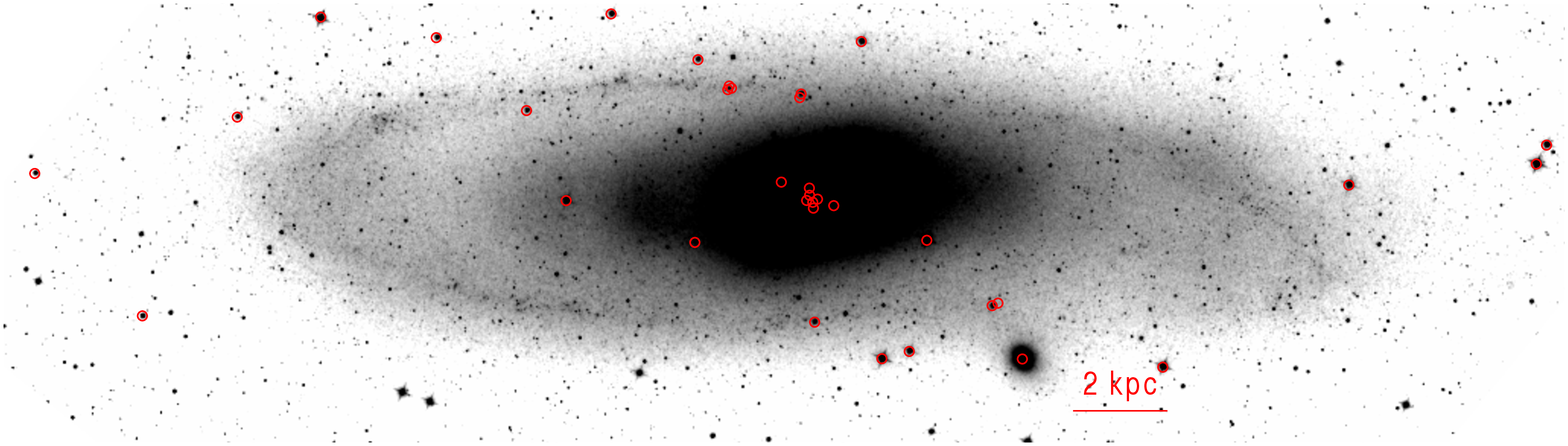}
\includegraphics[width=0.95\textwidth,trim=0 0 0 0,clip=true]{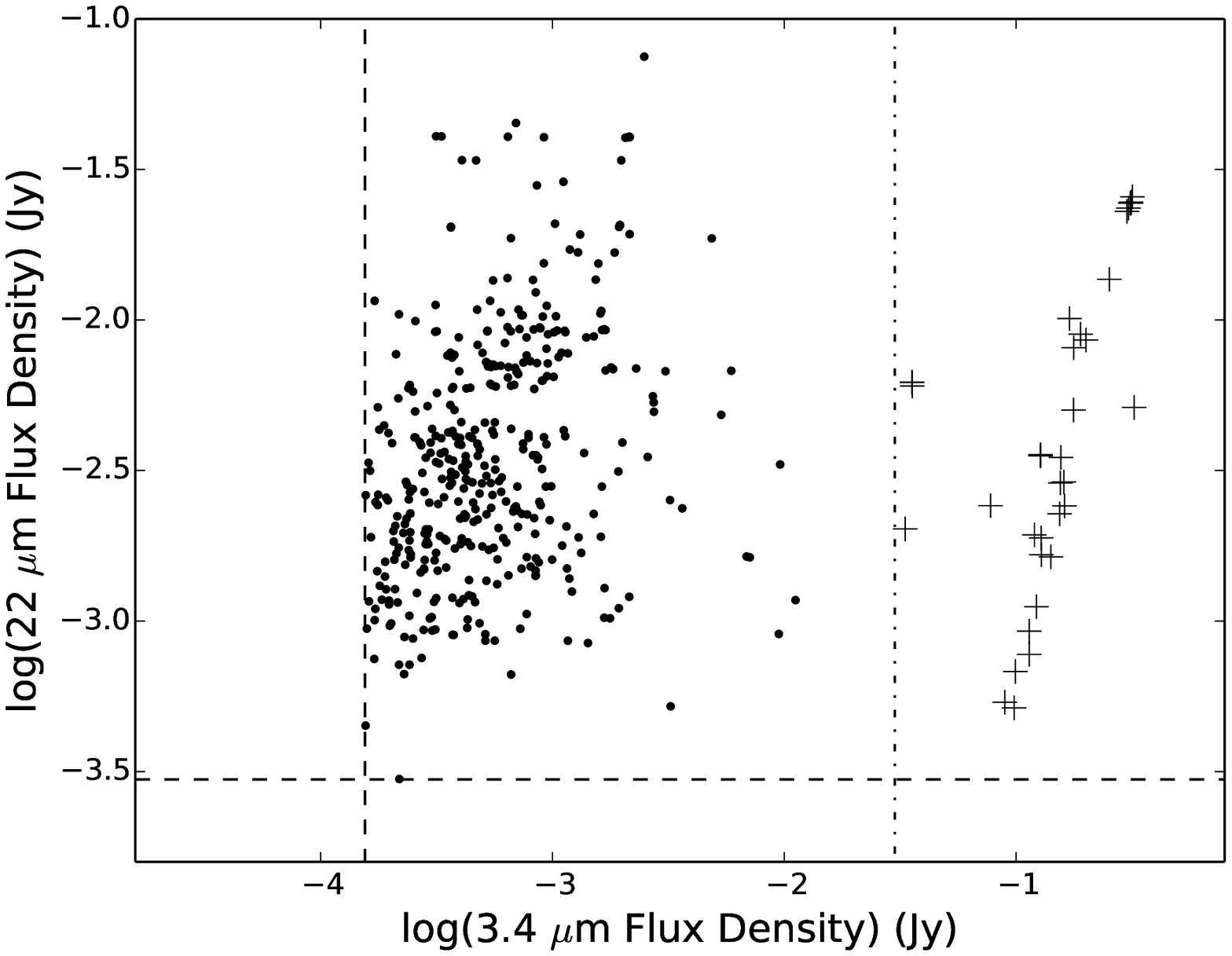}
\caption{Identification of stellar interlopers in the star-forming regions sample. Upper panel: 3.4 $\mu$m WISE image of M31. The red circles show the giant star candidates identified in our sample. 
Plot:  22~$\mu$m vs. 3.4~$\mu$m flux density for the complete survey sample. Crosses indicate objects identified as giant stars, and filled circles indicate the rest of the sample. The dashed lines show the 3$\sigma$ detection limit at 3.4 $\mu$m and the 5$\sigma$ detection limit at 22 $\mu$m. The vertical dashed-dotted line at 0.03 Jy separates the candidate giant stars from the rest of the sample.}
\label{fig: wise_noHII}
\end{figure*}

Figure \ref{fig: wise_noHII} shows the 22~$\mu$m vs. 3.4~$\mu$m emission for the sample. The aperture radius of 5.5\arcsec\ was used to measure the flux density at 3.4 and 22~$\mu$m. We found that for 35 regions there is a
clear separation from the rest of the sample, suggesting that these are giant stars.  We impose a 3.4~$\mu$m cut
at 0.03 Jy to separate giant stars from star-forming regions:  giant stars form the more luminous 
population at 3.4~$\mu$m.  
Table \ref{tab:removed} lists these 35 sources identified as giant stars, including their coordinates. 
We removed these giant stars from the sample, and after removing the nine PNe as well, the final source list includes 462 objects (five of which are maser sources) that are likely to be star-forming regions.

\floattable
\begin{deluxetable}{cc|cc}
\tabletypesize{\scriptsize}
\tablecaption{Planetary Nebulae and Giant Stars\label{tab:removed}}
\tablewidth{0pt}
\tablehead{
\colhead{Object (J2000)} &  \colhead{Classification} & \colhead{Object (J2000)} &  \colhead{Classification} 
}
\startdata
003838.7+402613.5   & Star &
003849.2+402551.7   & Star \\
003950.9+402252.1   & PNe  &
003954.4+403820.4   & Star \\
004040.4+402709.8   & PNe & 
004129.8+412211.1   & Star \\
004201.5+404115.7   & Star &
004208.9+412329.8   & Star \\
004210.7+412322.3   & Star &
004226.1+410548.2   & Star \\
004227.9+413258.5   & Star &
004228.1+405657.7   & Star \\
004228.3+412911.4   & Star &
004228.4+412852.4   & Star \\
004230.1+412904.0   & Star &
004230.9+405714.6   & Star \\
004237.4+414158.3   & Star &
004241.7+411435.0   & Star \\
004241.9+405155.2   & Star &
004242.6+411722.5   & Star \\
004244.4+411608.5   & Star &
004245.3+411656.9   & Star \\
004247.0+411618.4   & Star &
004248.2+411651.7   & Star \\
004249.1+411554.6   & Star &
004249.1+411945.9   & Star \\
004310.0+413751.6   & PNe &
004314.2+410033.9   & Star \\
004325.6+410206.4   & Star &
004329.2+414848.0   & PNe  \\
004332.5+410907.0   & Star &
004339.4+412229.2   & Star \\
004341.5+414224.3   & Star &
004341.7+415313.0   & Star \\
004355.8+411211.6   & PNe &
004403.9+413414.8   & Star \\
004410.5+420247.5   & Star &
004433.8+415249.7   & PNe \\ 
004435.6+415606.9   & PNe &
004515.9+420254.4   & Star \\
004540.0+415510.2   & PNe & 
004641.6+421156.2   & PNe \\  
004642.6+421406.8   & Star &
004703.1+415755.4   & Star \\
\enddata
\end{deluxetable}

\subsection{The Detection Sample}\label{sec:detection}

We construct a ``detection'' sample of 320 sources in the non-maser sample that have measured values for all properties for each object.  
The 24~$\mu$m flux densities and SFR were measured for 457 non-maser sources. We measure flux densities at 70~$\mu$m above 0.05~Jy (3$\sigma$) for 389 sources in the non-maser sample. At 160~$\mu$m we obtain flux densities for 447 non-maser sources above 0.06~Jy (3$\sigma$).  There are 346 H$\alpha$ counterparts for the non-maser sample. There are three sources in the non-maser sample with no temperature in the dust temperature map. The intersection of these sets includes 320 regions.

We measure all values for the water maser sample, except for the H$\alpha$ flux for the source 004409.4$+$411856.3. Since other properties for 004409.4$+$411856.3 were measured (Table \ref{tab:maser}), we include this source in the analysis. We only removed this object from the water maser sample where H$\alpha$ flux was involved in the statistics. This includes computing the correlation between H$\alpha$ and other variables and the Kolmogronov-Smirnov (K-S) test on H$\alpha$ in the maser and non-maser samples. Below we describe the results of statistical analyses we performed on the water masers and the "detection'' sample of non-maser regions.

\section{Results}\label{sec:results}

\begin{figure*}
\includegraphics[width=1.0\textwidth, trim=0 0 0 0,clip=true]{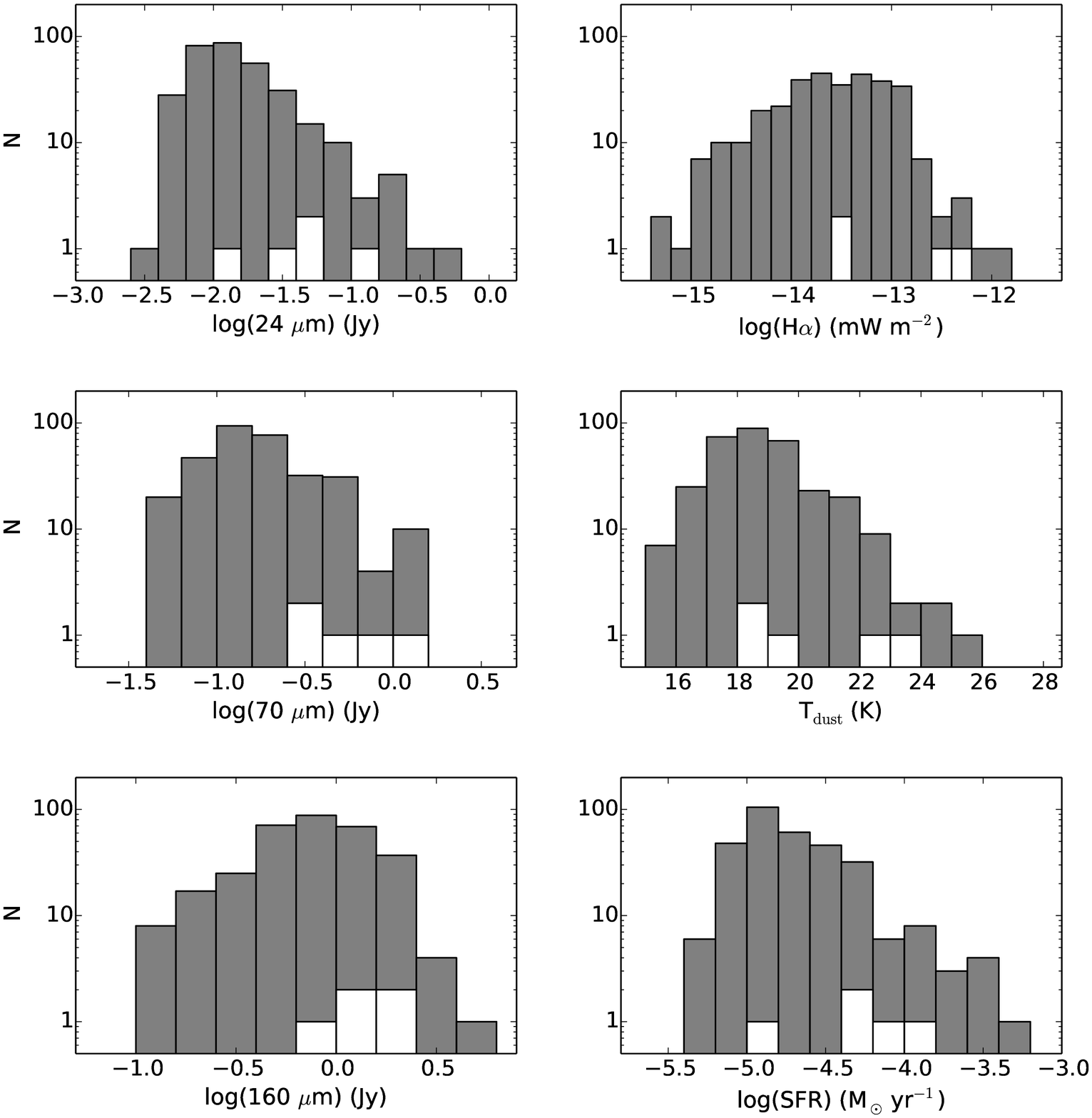}
\caption{Distribution of the 24~$\mu$m, 70~$\mu$m, and 160~$\mu$m flux densities, the H$\alpha$ flux, the dust temperature, and the star formation rate for the water maser and non-maser objects. White bars show the water maser hosts and the black bars indicate the non-masers.
}
\label{fig:hist}
\end{figure*}
 
Figure \ref{fig:hist} shows the distribution of the 24 $\mu$m, 70 $\mu$m, and 160 $\mu$m flux densities, the H$\alpha$ flux, the dust temperature, and the star formation rate for the water maser hosts and the non-maser sources in the detection sample, and Figure \ref{fig:regression} shows pairwise scatter plots in these quantities.   All values used in the following analysis are tabulated in Tables \ref{tab:maser}
and \ref{tab: photometry}.

\section{Analysis}\label{sec:analysis}
The GBT water maser survey of 506 luminous compact 24~$\mu$m regions produced five water maser detections. 
Sensitivity limitations of the observations ($\sim$10 mJy at 3$\sigma$) and imperfections in the selection criteria (44 objects
were subsequently omitted as likely PNe or giant stars) aside, the low detection rate of 1.1(0.5)\% \citep{darling2016} may indicate that water maser emission favors specific physical conditions in star forming regions in M31.
To explore this possibility, we examine the optical, FIR, dust temperature, and star formation rate characteristics of the water maser and non-maser regions used in the GBT survey.
Using two sample K-S tests, correlation statistics, principal component analysis, and survival analysis, 
we compare the properties of the water maser and non-maser samples.  We then examine the FIR-H$_2$O maser
luminosity relation and the role of star formation in the maser detection rate.

\floattable
\begin{deluxetable}{ccccccc}
\tabletypesize{\footnotesize}
\tablecaption{K-S Test of  Water Maser Regions and Non-Maser Regions\label{tab: K-S}}
\tablewidth{0pt}
\tablehead{
\colhead{ } & \colhead{T$_{\rm dust}$} & \colhead{log(SFR)}  & \colhead{log(24~$\mu$m)} & \colhead{log(70~$\mu$m)} & \colhead{log(160~$\mu$m)} & \colhead{log(H$\alpha$)} 
}
\startdata
p-value & 0.19 & 0.013 & 0.013 & 0.00075 & 0.041 & 0.11 \\
\enddata
\end{deluxetable}

\subsection{One-Parameter Tests}

\subsubsection{K-S Tests}\label{K-S}

We performed non-parametric two-sample K-S tests to examine the differences in properties of the water maser and non-maser regions. The p-values of these tests are listed in Table \ref{tab: K-S}.  The star formation rate and the 24~$\mu$m, 70~$\mu$m, and 160 $\mu$m flux densities have a p-value less than 0.05, suggesting significant differences between the maser-emitting regions and those that do not show maser emission.  The dust temperature and 
H$\alpha$ emission do not show significant differences.  Note that the maser sample is small, and that one
of the masers, 004409.4$+$411856.3, is excluded from the H$\alpha$ statistic.  

\subsubsection{Survival Analysis}\label{sec:survival}

As discussed section \ref{sec:detection}, some values are listed as upper limits in Tables \ref{tab:maser} and  \ref{tab: photometry}, and these objects were excluded from the ``detection'' sample analyzed above.  
Altenatively, upper limits can be included in a survival analysis.  
We performed two-sample tests to study the difference between the measured (censored) properties of the 
water maser and non-maser regions using 
the survival analysis package Nondetections and Data Analysis for environmental data (NADA)\footnote{\url{http://cran.r-project.org/web/packages/NADA/NADA.pdf}} implemented in 
R\footnote{\url{https://www.R-project.org}}.  The NADA package has been shown to give results appropriate for astronomical data that includes non-detections \citep[][]{feigelson2013}.
We used the NADA ``cendiff'' routine and performed the  Peto $\&$ Peto two-sample test, which is an appropriate treatment for left-censored data that include upper limits. 

There are no dust temperatures for five sources in the non-maser sample, and since one cannot place upper limits on the dust temperature in these cases, these sources were omitted from the survival analysis. 
Table \ref{tab:two sample_survival} shows the results of the two-sample test performed on the water maser and non-maser samples. The p-values indicate the probability that the two samples are drawn from the same distribution. The p-values for the dust temperature, the SFR, and the 24~$\mu$m, 70~$\mu$m, and 160~$\mu$m
flux densities show that the water maser and non-maser samples are not mutually consistent.  The p-value for H$\alpha$ flux, however, indicates that the water maser and non-maser samples are indistinguishable.
These results are similar to the K-S test results for the ``detection'' sample (Section \ref{K-S}) for the 
SFR and the 24~$\mu$m, 70~$\mu$m, and 160 $\mu$m flux densities, suggesting that the censorship on the 
samples did not significantly affect inferences made based only on the ``detection'' sample.  

\floattable
\begin{deluxetable}{ccccccc}
\tabletypesize{\footnotesize}
\tablecaption{Survival Analysis of Water Maser and Non-Maser Regions\label{tab:two sample_survival}}
\tablewidth{0pt}
\tablehead{
\colhead{ } & \colhead{T$_{\rm dust}$} & \colhead{log(SFR)}  & \colhead{log(24~$\mu$m)} & \colhead{log(70~$\mu$m)} & \colhead{log(160~$\mu$m)} & \colhead{log(H$\alpha$)} 
}
\startdata
p-value & 0.0086 & $3.5 \times 10^{-6}$ &  $3.6 \times 10^{-7}$& $1.0 \times 10^{-11}$ & $4.2 \times 10^{-5}$ & $0.055 $  \\
\enddata
\end{deluxetable}

\subsection{Two-Parameter Tests} \label{correlation}

We performed a correlation analysis of the properties of the water maser and non-maser regions separately.
Table \ref{tab:maser hosts correlation} lists the Pearson correlation coefficients for the water maser host properties, 
including the water maser line-integrated flux densities,
and Table \ref{tab:maser hosts p-value} lists the p-values associated with Pearson correlation coefficients 
(the probability that the correlation occurs by chance).  Again, the H$\alpha$ flux of 004409.4$+$411856.3
is omitted from the analysis, but all other properties of this maser region are included.  

Significant correlation exists between SFR, 24 $\mu$m, and 70 $\mu$m emission because these quantities
are all driven by (or tautologically are) star formation.  p-values are slightly larger than 0.05 for the correlations
between SFR and other star formation-driven quantities, such as 160 $\mu$m and H$\alpha$ emission, most
likely due to the very small sample size (see non-maser regions, below).  Dust temperature also shows 
a significant correlation with H$\alpha$.  Water maser emission (flux) notably shows no significant correlation 
with any other measured property (but see Section \ref{subsec:FIR-H2o}), which is not surprising given the very small-scale emission regions 
of water masers that amplify local conditions.  The bulk properties of star-forming regions, however, may 
still be predictive of the formation of water masers, even if one cannot predict the luminosity of the emitted 
masers (Section \ref{subsec:detection}).

Table \ref{tab:non-maser hosts} lists the Pearson correlation coefficients for the non-maser sample, and 
Table \ref{tab:non-maser hosts p-value} lists the associated p-values.  All quantities are significantly correlated among the non-maser sample because all of the properties under study are related to star formation in the
``detection'' sample.  

\begin{figure*}[t]
\includegraphics[width=1.0\textwidth,trim=0 0 0 0,clip=true]{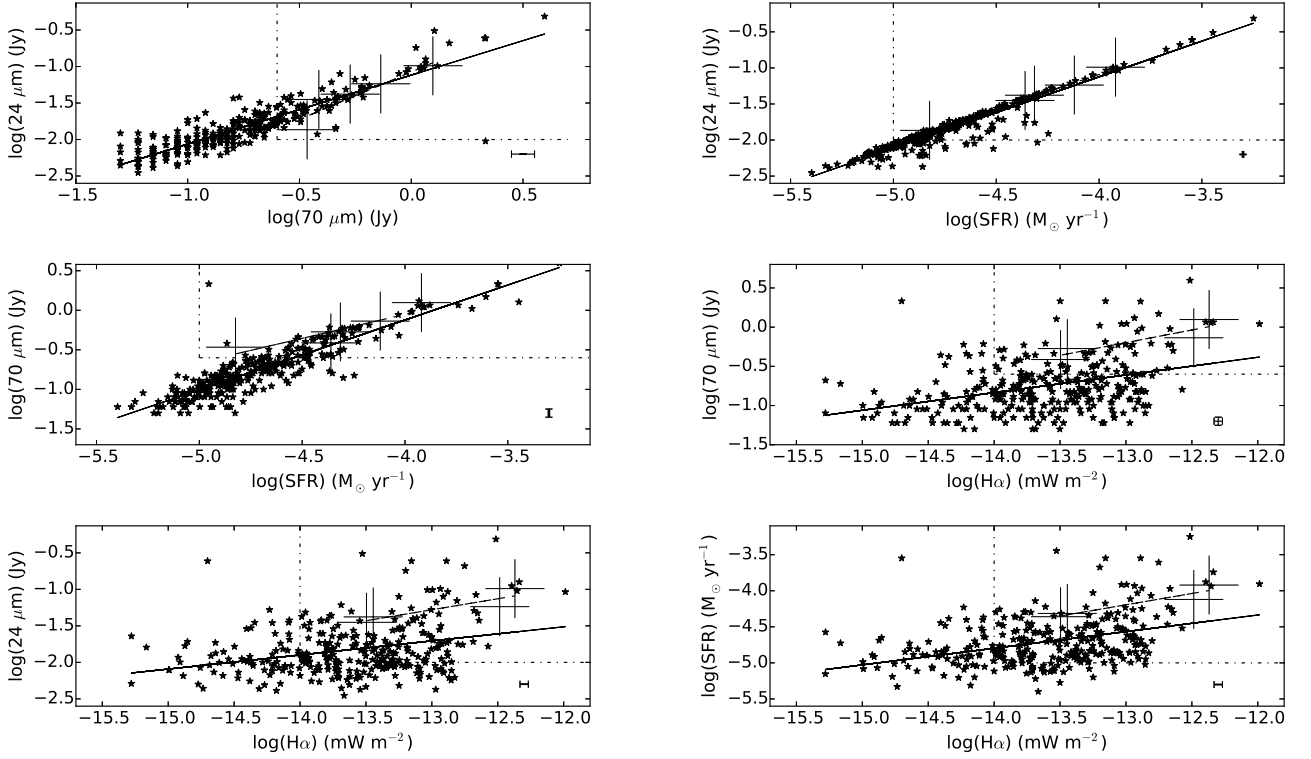} 
\caption{ Relationships between the properties of the water maser and non-maser regions. Crosses indicate the water masers (enlarged for clarity), and stars indicate the non-masers. The dash-dotted lines show the regions of parameter space where water masers arise and where new water masers may be found (see text). 
Representative 1$\sigma$ error bars are shown in the lower right of each panel.  }
\label{fig:regression}
\end{figure*}

Figure \ref{fig:regression} shows scatter plots of various pairs of properties for both maser and non-maser samples,
focusing on those that show correlation in the water maser-emitting regions.  
The dashed-dotted lines indicate the regions of parameter space where masers are found, and where
future surveys might concentrate.   The loci are 
$\log(24\mu{\rm m}) > -2.0$, 
$\log(70\mu{\rm m}) > -0.6$, 
$\log({\rm SFR}) > -5.0$, and
$\log({\rm H}\alpha ) > -14.0$, in the units listed in Tables \ref{tab:maser} and \ref{tab: photometry}.   
The intersection of all of these limits reduces the sample to 70 sources, including four masers, yielding a detection 
rate of 5.7(2.8)\%.

\floattable 
\begin{deluxetable}{lccccccc}
\tabletypesize{\footnotesize}
\tablecaption{Pearson Correlation Coefficients for the Water Maser-Emitting Regions in M31\label{tab:maser hosts correlation}}
\tablewidth{0pt}
\tablehead{
\colhead{ } & \colhead{T$_{\rm dust}$} & \colhead {log(SFR)} & \colhead{log(24~$\mu$m)} & \colhead{log(70~$\mu$m)} & \colhead{log(160~$\mu$m)} &  \colhead{log(H$\alpha$)} & \colhead{log(H$_2$O)}
}
\startdata
T$_{\rm dust}$     &  1.00      &    0.78   &    0.76     &     0.82      &    0.36     &     0.96      &    0.54  \\
log(SFR)           &          	&    1.00   &    1.00     &     0.90      &    0.84     &     0.94      &    0.23  \\
log(24 $\mu$m)     &  		&     	    &	 1.00     &     0.92      &    0.87     &     0.90      &    0.18  \\
log(70 $\mu$m)     &  		&	    &	   	  &	 1.00     &    0.76     &     0.90      &    0.21  \\
log(160 $\mu$)     &  		&	    &		  &	 	  &    1.00     &     0.23      &   -0.19  \\
log(H$\alpha$)     &  		&	    &		  &		  &             &     1.00      &    0.46  \\
log(H$_2$O)        &            &           &             &               &             &               &      1.00    \\
\enddata
\end{deluxetable}

\floattable 
\begin{deluxetable}{ccccccccc}
\tabletypesize{\footnotesize}
\tablecaption{Pearson Correlation Coefficient p-values for the Water Maser-Emitting Regions\label{tab:maser hosts p-value}}
\tablewidth{0pt}
\tablehead{
\colhead{ } & \colhead{T$_{\rm dust}$} & \colhead {log(SFR)	} & \colhead{log(24~$\mu$m)} & \colhead{log(70~$\mu$m)} & \colhead{log(160~$\mu$m)} &  \colhead{log(H$\alpha$)} & \colhead{log(H$_2$O)}
}
\startdata
T$_{\rm dust}$     &	 &  0.12&  0.14 &  0.09 &  0.56 &  0.04 &  0.35  \\
log(SFR)           &     &  	&  0.00 &  0.04 &  0.07 &  0.06 &  0.71	 \\
log(24 $\mu$m)     &     & 	& 	&  0.03 &  0.05 &  0.10 &  0.77	 \\
log(70 $\mu$m)     &     &	&	& 	&  0.13 &  0.10 &  0.73	 \\
log(160 $\mu$)     &     & 	&	&	&       &  0.77 &  0.76	 \\
log(H$\alpha$)     &     & 	&	&	&       &	&  0.54	 \\
log(H$_2$O)        &     &      &       &       &       &       &        
\enddata
\end{deluxetable}

\floattable 
\begin{deluxetable}{ccccccc}
\tabletypesize{\footnotesize}
\tablecaption{Pearson Correlation Coefficients for the Non-Maser Sample in M31\label{tab:non-maser hosts}}
\tablewidth{0pt}
\tablehead{
\colhead{ } & \colhead{T$_{\rm dust}$} & \colhead{log(SFR)} & \colhead{log(24 $\mu$m)} & \colhead{log(70 $\mu$m)} & \colhead{log(160 $\mu$m)} & \colhead{log(H$\alpha$)} 
}
\startdata
T$_{\rm dust}$      &  1.00 &  0.42  & 0.37   &   0.43      &    0.15     &  0.29   \\
log(SFR)         &   &  1.00 & 0.96   &   0.89      &    0.62     &  0.38   \\
log(24 $\mu$m)    &   &        &  1.00  &   0.90      &    0.64     &  0.31   \\
log(70 $\mu$m)    &   &        &        &       1.00  &    0.69     &  0.38   \\
log(160 $\mu$m)   &   &        &        &             &    1.00   &  0.21   \\
log(H$\alpha$)   &   &        &        &             &             &   1.00  \\
\enddata
\end{deluxetable}

\floattable
\begin{deluxetable}{ccccccc}
\tabletypesize{\footnotesize}
\tablecaption{Pearson Correlation Coefficient p-values for the Non-Maser Sample\label{tab:non-maser hosts p-value}}
\tablewidth{0pt}
\tablehead{
\colhead{ } & \colhead{T$_{\rm dust}$} & \colhead{log(SFR)} & \colhead{log(24 $\mu$m)} & \colhead{log(70 $\mu$m)} & \colhead{log(160 $\mu$m)} & \colhead{log(H$\alpha$)} 
}
\startdata
T$_{\rm dust}$    & & $   3.1\times 10^{-15}  $ & $    6.4 \times 10^{-12}  $ & $   7.8\times 10^{-16}    $ & $ 5.7 \times 10^{-3}    $ & $1.2 \times 10^{-7}     $\\
log(SFR)             &&  & $    2.1\times 10^{-175}   $ & $  3.0\times 10^{-113}    $ & $ 7.2 \times 10^{-35}    $ & $9.7 \times 10^{-13}  	  $ \\
log(24 $\mu$m)        && $                       $ & $                      $ & $  8.5 \times 10^{-119}   $ & $  9.4 \times 10^{-39}   $ & $ 1.2 \times 10^{-8}    $	\\
log(70 $\mu$m)        && $                       $ & $                          $ & $                  $ & $    7.8 \times 10^{-47} $ & $   1.7 \times 10^{-12}   $	\\
log(160 $\mu$m)        && $                       $ & $                          $ & $                         $ & $                 $ & $      1.9 \times 10^{-4}$  \\
log(H$\alpha$)       && $                       $ & $                          $ & $                         $ & $                        $ & $                    $ \\
\enddata
\end{deluxetable}

\subsection{FIR-H$_2$O Maser Luminosity Relation}\label{subsec:FIR-H2o}
Previous studies have shown that water masers in the Galaxy are associated with compact bright FIR sources. \cite{jaffe1981} found that H$_2$O maser emission in star-forming regions is associated with 50--100\% of the bright Galactic FIR sources and that H$_2$O maser luminosities are correlated with FIR luminosities. 
\cite{felli1992} found a similar correlation between water maser luminosity and FIR luminosity in Galactic 
star-forming regions.
Extragalactic water masers show a rough correlation between the water maser luminosity and FIR luminosity \citep[][]{henkel2005, castangia2008}, although water megamasers and kilomasers appear to follow different correlations: for a given FIR luminosity, water kilomasers are sub-luminous compared to masers emitted from 
Galactic star-forming regions, and water megamasers appear to be slightly over-luminous. 

\begin{figure}
\center
\includegraphics[width=0.5\textwidth]{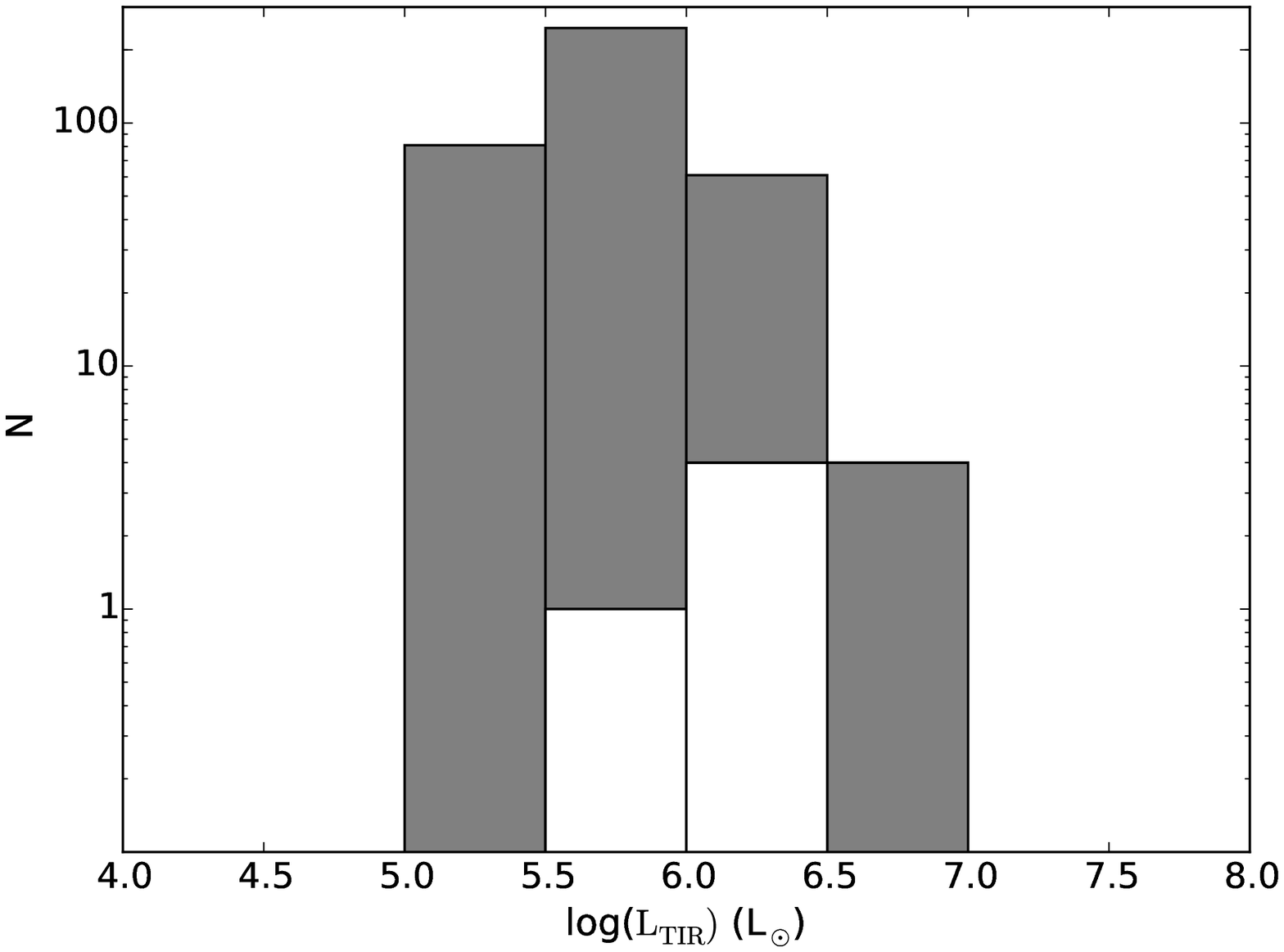}  
\caption{Distribution of the TIR luminosities of the M31 survey for the water masers (clear bins) and non-masers (shaded bins)  
listed in Tables \ref{tab:maser} and  \ref{tab: photometry}. }
\label{fig:Lbol_31}
\end{figure}

Since the M31 water masers in principle represent analogs to the high end of the Galactic water maser luminosity
distribution, it is useful to compare the FIR-H$_2$O luminosity relation for water masers in M31 to Galactic and Extragalactic water masers.  The infrared luminosity of star-forming regions in M31 can be calculated using Equation 4 in \citet{dale2002} that describes the total infrared (TIR) luminosities of galaxies in the range 3--1100 $\mu$m:
\begin{equation}\label{eq:TIR}
 L_{ TIR} = \zeta_1\, \nu L_\nu(24\ \mu{\rm m}) +\zeta_2\, \nu L_\nu(70\ \mu{\rm m}) \\
+ \zeta_3\, \nu L_\nu(160\ \mu{\rm m}),
\end{equation}
where $[\zeta_1,\ \zeta_2,\ \zeta_3] $= [1.559, 0.7686, 1.347] for $z=0$, and $\nu$ is the frequency in Hz. \cite{dale2002} explain that Equation \ref{eq:TIR} matches the model bolometric infrared luminosity to better than 1$\%$ at $z=0$. It is therefore reasonable to assume that the TIR and bolometric luminosities of the water maser and non-maser samples can be obtained from Equation \ref{eq:TIR}. 
We convert the flux density values at 24 $\mu$m, 70 $\mu$m, and 160 $\mu$m in Tables \ref{tab:maser}
and \ref{tab: photometry} to specific luminosities assuming a distance of 780 kpc ($ L_\nu =   4 \pi  D^2 S_\nu$) and
then calculate $L_{TIR}$.   Tables \ref{tab:maser} and \ref{tab: photometry} show the TIR  luminosities for the 
maser and non-maser samples, respectively (for the non-maser sample, $L_{TIR}$ is only calculated for the 
387 sources that are detected in all three FIR bands).  
Figure \ref{fig:Lbol_31} shows the distribution of the TIR luminosity of the water maser and non-maser sources in the M31 survey. 
Table \ref{tab:maser} also lists the isotropic H$_2$O maser luminosities obtained from \cite{darling2011}.

\begin{figure}
\center
\includegraphics[width=0.48\textwidth,trim=0 0 0 0,clip=true]{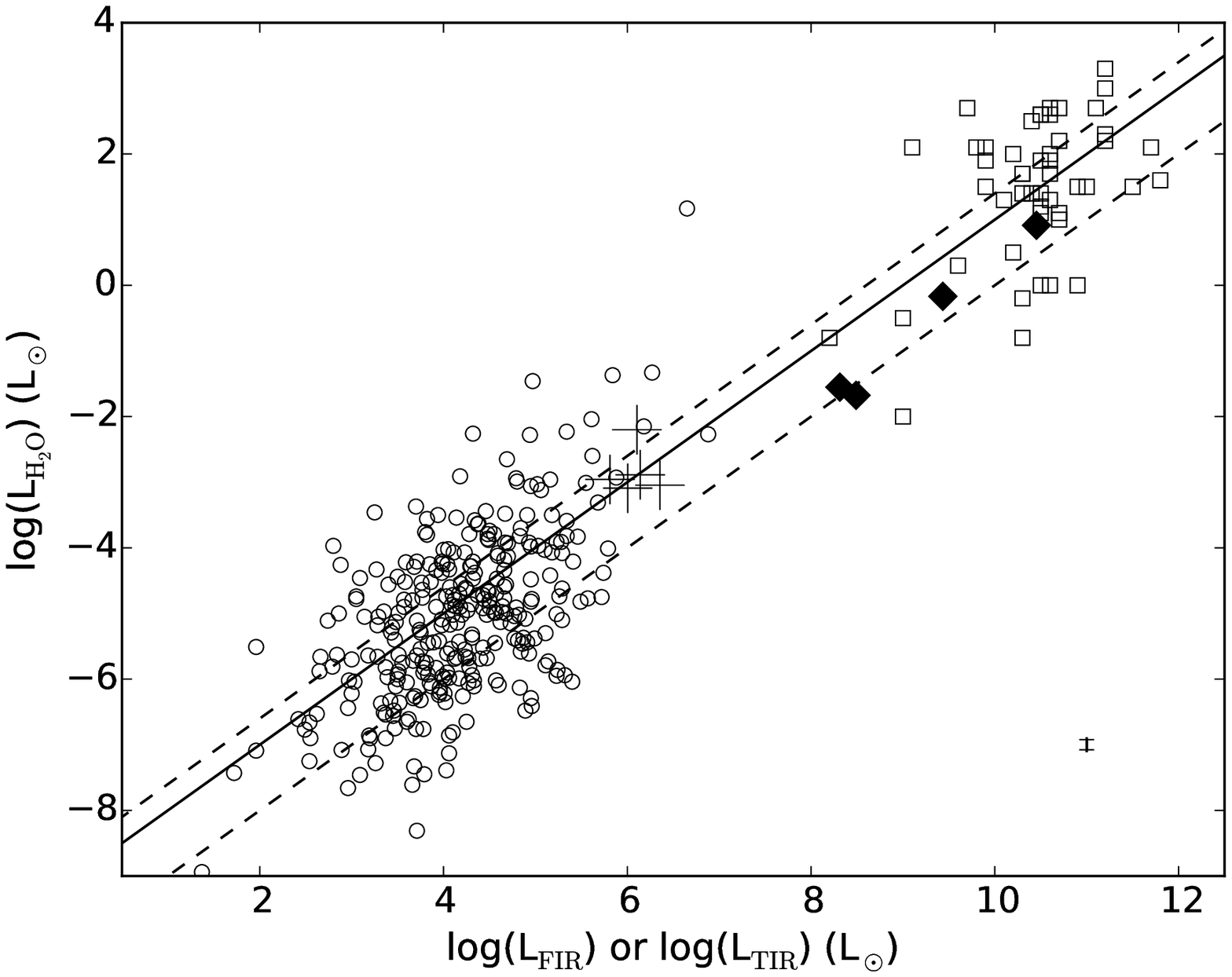}
\caption{ H$_2$O vs. FIR (or TIR) luminosity of Galactic and extragalactic water masers. The solid line shows the relation for Galactic star-forming regions \citep[][]{jaffe1981}, and the dashed lines show the relation for megamasers and kilomasers \citep[][]{henkel2005,castangia2008}. Crosses mark the water masers in M31, the open squares indicate extragalactic H$_2$O megamasers and kilomasers obtained from Table 4 of \cite{henkel2005} and references therein,  filled diamonds indicate kilomasers associated with star formation in nearby galaxies \citep[][]{darling2008}, and open circles mark FIR luminosities and H$_2$O maser luminosities for Galactic water masers obtained from \cite{urquhart2011}. Representative error bars indicate 1$\sigma$ uncertainties associated with the water masers in M31. }
\label{fig:fir_h2o}
\end{figure}

Figure \ref{fig:fir_h2o} shows the H$_2$O-FIR (or -TIR) luminosity relation for Galactic and extragalactic water masers. 
The water masers in M31 are clearly consistent with the Galactic relation obtained by \citet{jaffe1981} and overlap
the high-luminosity tail of the Galactic distribution.  
This result suggests two things:  (1) the M31 water masers do indeed seem to be analogs to the high-luminosity tail 
of the Galactic water maser distribution, suggesting that what is known about Galactic water masers can be 
applied to those in M31, and (2) a more sensitive survey of IR-luminous regions in M31 is likely 
to detect more water masers.

\subsection{The Maser Detection Rate}\label{subsec:detection}

The luminosity of the water masers in M31 seems to follow the same relationship with the FIR luminosity as do 
masers in the Galaxy and in other galaxies, but is the detection rate of water masers in line with the Galactic 
rate, scaled to the luminosity sensitivity of the GBT water maser survey?  

\begin{figure}
\center
\includegraphics[width=0.5\textwidth]{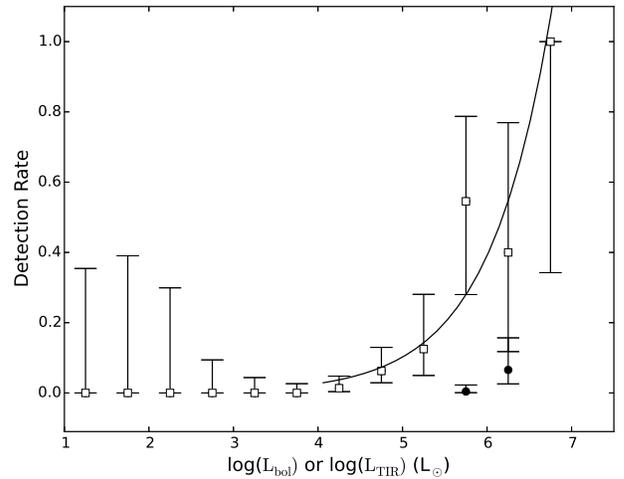}  
\caption{Water maser detection rate versus bolometric (or TIR) luminosity. Empty squares indicate the detection rate of the Galactic \cite{urquhart2011} sample, adjusted to the luminosity sensitivity of the GBT M31 water maser survey (4.4$\times$10$^{-4}$ $\rm L_\odot$). Filled circles indicate the detection rate of the GBT M31 water maser survey. 
The solid line depicts a least-squares fit to the non-zero \citet{urquhart2011} points.  Error bars indicate the 95$\%$ confidence interval.}
\label{fig:urquhart}
\end{figure}

In a survey of massive young stellar objects (MYSOs), compact HII regions, and ultra compact HII regions selected from the Red MSX Survey (RMS), \cite{urquhart2011} obtained an overall detection rate of $\sim 50\%$, but
demonstrated a strong correlation between the H$_2$O maser detection rate and the bolometric luminosity.  
In order to compare this result to the M31 maser detection rate, we restrict the \citet{urquhart2011} 
sample to the GBT luminosity sensitivity, $L_{H_{2}O} \ge 4.4 \times 10^{-4} \ L_{\odot}$, and re-calculate the
detection statistics as a function of bolometric luminosity.  Figure \ref{fig:urquhart} shows this censored Galactic detection rate that can be directly compared to the M31 detection rate.  

Due to the small sample size and the fact that the detection rates for some luminosity bins correspond to zero or one, the standard binomial confidence interval 
that relies on the central limit theorem is not applicable. Instead we estimate the error bars using Wilson's score interval
\citep{wilson1927}:
\begin{equation}\label{eq:wilson}
   \frac{1}{1 + z^{2}/n}\left [{\hat{p} + \frac{z^{2}}{2n} \pm z \sqrt{\frac{\hat{p}(1-\hat{p})}{n} + \frac{z^{2}}{4n^{2}}}}\  \right]\ ,
\end{equation}
where $z$ indicates the $ 1 - \alpha/2$ percentile of a standard normal distribution, $\hat{p}$ is the estimated 
detection rate, and $n$ is the number of samples. For a 95$\%$ confidence interval, $1 - \alpha/2 = 0.975$ and z = 1.96. An unconstrained least-squares fit to the non-zero data points for RMS sources obtains the following relationship:
\begin{equation}
 \log\left( {\rm Detection\ Rate} \right)= (0.58 \pm 0.02) \times \log L_{bol} - (3.9 \pm 0.49), 
\end{equation}
which is plotted in Figure \ref{fig:urquhart}.

The filled circles in Figure \ref{fig:urquhart} indicate the detection rate for the GBT M31 water maser survey.
The detection rates in the two luminosity bins are $0.004^{+0.019}_{-0.003}$ ($5.5 < \log L_{TIR}/L_{\odot} < 6.0$) and $0.066^{+0.091}_{-0.040}$ ($6.0 < \log L_{TIR}/L_{\odot} < 6.5$). The upper and lower bounds indicate the 95$\%$ confidence interval.
The water maser detection rates for the same luminosity bins in the censored \cite{urquhart2011} sample are $0.55^{+0.24}_{-0.26}$ and $0.40^{+0.37}_{-0.28}$, respectively. The number of water masers in these luminosity bins is 11 and 5, whereas the number in the corresponding M31 bins is 1 and 4.  Due to the small number of detected masers, these detection rates have
large uncertainties, and the higher luminosity bin rates are statistically consistent between the two surveys (we omit from this statement the M31 bin that has a single maser).  The M31 water maser detection rate thus does not significantly differ from the Galactic
rate, although this is a statement more about poor statistics rather than about the nature of the masers in M31.

\section{Discussion}\label{sec:discussion}

Water masers associated with star forming regions are known to trace shock regions in the outflows from both high mass and low mass young stellar objects \citep[YSOs,][]{honma2005, goddi2006, moscadelli2006}. VLBI observations often indicate water masers trace disks and outflows \citep[e.g.][]{seth2002}. The most powerful masers in the Galaxy are associated with water masers in high-mass star-forming regions and flux densities of more than $10^4$~Jy have been observed during maser flares (e.g., W49N) \citep[e.g.,][]{liljestrom1989}.  The water masers in M31, therefore, may show both offset velocities
and significant variability, both of which may frustrate detection and proper motion studies.

\subsection{Water Maser Surveys in the Galaxy}
Water masers in the Galaxy have been observed toward ultra-compact HII (UCHII) regions and bright FIR sources. UCHII regions are small photoionized nebulae that are associated with earliest stages of massive star formation.
\cite{churchwell1990} surveyed a sample of bright IRAS FIR color-selected UCHII regions and detected water masers brighter than 0.4 Jy toward $\sim 67\%$ of the sources. Several other surveys have been performed toward known UCHII regions in the Galaxy.  For example, \citet{palla1991} obtained an overall detection rate of 17\% using a 5 Jy detection limit (3$\sigma$), \citet{codella1995} detected only 7\% with a 6 Jy limit (3$\sigma$), and \citet{kurtz2005} detected 55\% using a 0.34 Jy sensitivity (3$\sigma$).
The variable detection rates are due to selection criteria and sensitivity limitations 
\citep[see][for a detailed discussion]{kurtz2005}.

A catalog of 300 extended green objects (EGOs) were identified from the Spitzer Galactic Legacy Infrared Mid-Plane Survey Extraordinaire (GLIMPSE) project \citep[][]{cyganowski2008}. EGOs may represent massive young stellar objects still embedded in their infalling envelopes. They were identified based on their extended 4.5~$\mu$m (``green'') emission that likely traces shocked molecular gas from protostellar outflows. A survey of water masers in 94 northern EGOs showed a water maser detection rate of $68\%$ for a median rms sensitivity of $\sim$0.11 Jy \citep[][]{cyganowski2013}. 

\citet{walsh2011} performed a water maser survey of 100 deg$^{2}$ of the southern galactic plane using the Mopra Radio Telescope with rms noise of 1--2 Jy. They found 540 water masers, of which 340 are new detections. Based on the comparison of Galactic latitude distribution of the newly detected water masers with star-forming regions, they estimate at least 90$\%$ of the new detections originate in high-mass star-forming regions.

\subsection{Evolutionary Stage of Water Masers}\label{sec:evolutionary stage}

The evolutionary stage of water masers and which phase of massive star formation supports maser emission has been a topic of debate.
Recently, \citet{sanchez2013} studied  the water maser and 22 GHz continuum emission of 194 southern massive star forming regions. They classified MYSOs based on the evolutionary scheme proposed by \citet{molinari2008} into three types, in order of increasing age:

 \begin{itemize}
 \item Type 1: Millimeter only sources. They are mainly high-mass protostars embedded in dusty clumps
 \item Type 2: Millimeter plus infrared sources. These objects are mainly zero age main sequence (ZAMS) OB stars with compact HII regions, but still embedded in dusty clumps.
 \item Type 3: Infrared-only sources. These objects are more evolved ZAMS OB stars surrounded by remnants of their parental clouds and contain extended and less dense HII regions.
 \end{itemize}

Water masers were found to be mainly associated with Type 1 (13$\%$) and Type 2 (26$\%$) objects. Only 3\% of  Type 3 sources showed water masers. \citet{sanchez2013} indicate that their results are consistent with evolutionary schemes \citep[e.g.,][]{breen2010} where water masers appear at the early stage of massive star formation, co-exist with HII regions and disappear while HII regions are still observable. 

The bright 24 $\mu$m sources selected for the GBT M31 water maser survey belong to Type 2 and Type 3 classes. 
\citet{sanchez2013} obtained an overall H$_2$O maser detection rate of 12\% for Type 2 and Type 3 objects 
based on a single maser component sensitivity of $\sim$2~Jy (5$\sigma$).
Since 350 sources in the GBT survey are optically identified HII regions in the \citet{azimlu2011} H$\alpha$ catalog, 
the GBT survey selection method may include many objects that are least likely to produce water masers (Type 3 objects).
A more favorable means to detect 
water masers in M31 may be to select objects that show both mid-IR and mm continuum emission (Type 2 objects; see below).

Among the 457 non-maser sources, 346 objects have H$\alpha$ counterparts.  The 111 objects with no H$\alpha$ identification 
either have low H$\alpha$ luminosity because they are in an early stage of star formation or they are
highly obscured (or both).  Among the five masers, one shows no H$\alpha$ emission, so the fraction of masers without
H$\alpha$ emission is in agreement with the fraction among the non-masers:  $20^{+22}_{-12}$\% versus $24\pm2$\%, respectively.  

To identify possible maser sites that may be have been omitted from the GBT survey, 
we examined the 24~$\mu$m Spitzer map and the optically selected \citet[][]{azimlu2011} HII region catalog.
We find at most 10 sources that show a strong correspondence between H$\alpha$ and 24~$\mu$m 
emission that had not been selected for the GBT water maser survey. Assuming the most optimistic maser detection rate of 26$\%$ (Type 2 objects) in the \citet{sanchez2013} survey, and taking into account the poorer
luminosity sensitivity of the GBT survey, we would expect to find less than one additional maser in M31. This implies that we have likely detected the majority of water masers in M31 at the GBT survey sensitivity.

 In order to increase the number of water masers detected in M31, one would like to perform water maser surveys of bright millimeter sources. The GBT surveys \citet{darling2011,darling2016} targeted compact 24 $\mu$m sources in M31, which is associated with dusty molecular clouds and presumably star-forming regions. While this selection criterion includes a large fraction of active star forming regions in M31, it does not necessarily include all MYSOs embedded in dusty clumps that are only detectable at millimeter wavelengths (Type 1 objects) that show a 13\% detection rate \citep{sanchez2013}.
Currently, there is no published millimeter continuum map of M31 available for a water maser survey focused on the 
earliest stages of star formation.

\subsection{Comparison to Theory}

High resolution observations of water masers in star-forming regions indicate that water masers are associated with shocks \citep[e.g.][]{goddi2011,moscadelli2011}. \citet{hollenbach2013} present a theoretical model where 
water masers occur behind shocks with pre-shock densities in the range $\sim 10^{6}$--$10^{8}$ cm$^{-3}$. High velocity dissociative J shocks ($V_{s} \gtrsim 30$ km~s$^{-1}$) maintain a sufficient column of gas with temperatures ($\sim$300--400 K) heated by re-formation of H$_{2}$ to enable maser action. In this scenario a planar disk or slab with diameter $2\ell$ (along the line of sight) and thickness $d$ (perpendicular to the line of sight) maintains a large velocity coherent path length for maser action behind shock front. The aspect ratio of the maser geometry is defined as the 
radio of the path length to the maser spot size:  $ a = 2\ell / d$.  Using this model, \cite{hollenbach2013} relate the observed maser parameters including the brightness temperature, luminosity, and maser spot size to the physical properties of the shock region.  

For the five water maser complexes in M31, the size of maser spots and their brightness temperatures are unknown, 
and may remain unknown even with VLBI observations (the maser spot size is expected to be smaller than the 
best possible ground-based angular resolution).   Also, observational determination of the shock velocity is not straightforward: proper motion measurements are required to determine the shock velocity of the gas, but 
this does not necessarily directly relate to the shock velocity \citep[][]{hollenbach2013}. 
The maser luminosities, however, can be measured and compared to theory.

The observed luminosities of individual maser lines in M31 \citep[][]{darling2011} can be used to investigate
the properties of the shock regions. The luminosities of the water masers in M31 are in the range 
$3.2\times10^{-4}$~$ L_\odot$ to $1.9\times10^{-3}$~$ L_\odot$. 
\cite{hollenbach2013} show that the isotropic water maser luminosity of maser spots ranges from $3 \times 10^{-7}$~$ L_\odot$ to $10^{-5}$~$ L_\odot$ for  aspect ratio $a=10$ and  pre-shock densities of $\sim 10^6$--$10^8$~cm$^{-3}$. 
Since luminosity scales with aspect ratio as $a^3$, aspect ratios of $\sim$30--180 are required to achieve the observed isotropic luminosities of water masers in M31. Similarly, \cite{hollenbach2013} examine the brightest maser spot in W49N 
($ L_{ iso} = 0.08$~$ L_\odot$) and conclude that an aspect ratio of $a \ge 200$ is needed to produce the observed
isotropic luminosity.  Alternatively, high aspect ratios can arise from the alignment of two maser-producing regions
\citep{deguchi1989,elitzur1991}.
Clearly, the M31 water masers are exceptional, but they are consistent with the high-luminosity tail
of the Galactic distribution.

The M31 water maser detection rate is 1.1(0.5)\% \citep{darling2016}.  If all surveyed regions are producing masers, 
then the observed fraction $C$ of maser-emitting regions indicates the maser emission angle:  
$C \sim \sin \theta_{\rm em}$ \citep{hollenbach2013}.  
Since it is more likely that only a fraction of the observed regions are producing masers, this becomes a lower bound 
on the maser emission angle:  $\theta_{\rm em} \simeq 1/(2a) \gtrsim 0.6^\circ$.  This implies an aspect ratio $a \lesssim 45$, which is generally reasonable, but excludes the more extreme aspect ratios required by the high maser isotropic luminosities.  The resolution of this tension likely lies in the fact that we are not detecting individual maser spots in M31, 
and therefore the isotropic luminosity of the maser spots is lower than we have inferred from single dish observations.

\section{Conclusions}\label{sec:conclusions}

The multi-wavelength data used in this work enabled a comparative study of the properties of water maser-emitting regions and non-maser-emitting regions in M31.  An enhanced network of water masers in M31 would enable precise proper motion and the proper rotation measurements of M31, but there do not seem to be many additional water masers to be found in the star-forming regions of the galaxy at the current survey sensitivity.  We suggest three ways to detect
additional water masers in M31:

\medskip 

\noindent
1.  Re-observe the most luminous mid- or far-IR sources with higher sensitivity than was used by the GBT.  The known 
water masers in M31 represent the most luminous tail of the distribution, and improving a survey's sensitivity by a factor
of a few should produce an order of magnitude more maser detections.  The caveat with this approach is that these masers may not be bright enough for proper motion measurements using the sensitivity of current facilities.  

\medskip 

\noindent
2. Observe early-stage star-forming regions selected by mm continuum that have not been selected by their 24~$\mu$m 
emission.  The detection rate among such a sample will be low, but the mm continuum selection offers a means to detect
additional very luminous water masers that were missed in the GBT survey.

\medskip 

\noindent
3.  Re-observe the most luminous mid- or far-IR sources, and rely on maser variability for new detections.   Masers 
are highly variable and short-lived, so among a sample of $\sim$500 regions, newly luminous masers are a strong 
possibility over time baselines of 5--10 years.

In summary, this work demonstrated that:

\begin{itemize}
\item  Water masers are associated with the highest star formation rate regions in M31, and a good detection 
strategy is to focus on the most luminous regions in any star formation proxy (mid- and far-IR or H$\alpha$ 
luminosity). 

\item  The water masers in M31 are consistent with being analogs to water masers in Galactic star-forming regions
and represent the high-luminosity tail of a larger (and as yet undetected) population.  What is known about Galactic
water masers can probably be applied to the water masers in M31.  
\end{itemize}

\acknowledgments
The authors acknowledge support from the NSF grant AST-1109078.
We are grateful to Matthew Smith and George Ford from the Herschel Exploitation of Local Galaxy Andromeda (HELGA) team for providing access to Herschel SPIRE, dust temperature, and star formation rate maps of M31. We especially thank Bruno Altieri from European Space Agency (ESA) for re-processing the Herschel PACS map of M31.  The authors thank K.\ Gordon for the Spitzer map and the anonymous referee for helpful comments.  
This research made use of NASA's Astrophysics Data System Bibliographic Services, the NASA / IPAC Infrared Science Archive, 
and the NASA/IPAC Extragalactic Database
(NED) and uses observations made with the {\it Spitzer Space Telescope},
all of which are operated by the Jet Propulsion Laboratory,
California Institute of Technology, under a contract with NASA. 
This research made use of the observations made by the 
{\it Herschel Space Observatory}, an ESA space observatory with 
science instruments provided by European-led Principal Investigator 
consortia and with important participation from NASA. This publication makes use of data products from the Wide-field Infrared Survey Explorer, which is a joint project of the University of California, Los Angeles, and the Jet Propulsion Laboratory/California Institute of Technology, funded by the National Aeronautics and Space Administration. 
This research also made use of Montage, funded by NASA's Earth Science Technology Office, Computational Technologies Project, under Cooperative Agreement Number NCC5-626 between NASA and the California Institute of Technology. The code is maintained by the NASA/IPAC Infrared Science Archive.
Finally, this research made use of Astropy, a community-developed core Python package for Astronomy \citep[][]{astropy2013}.


\begin{thebibliography}{}

\bibitem[{{Astropy Collaboration} {et~al.}(2013){Astropy Collaboration},
  {Robitaille}, {Tollerud}, {Greenfield}, {Droettboom}, {Bray}, {Aldcroft},
  {Davis}, {Ginsburg}, {Price-Whelan}, {Kerzendorf}, {Conley}, {Crighton},
  {Barbary}, {Muna}, {Ferguson}, {Grollier}, {Parikh}, {Nair}, {Unther},
  {Deil}, {Woillez}, {Conseil}, {Kramer}, {Turner}, {Singer}, {Fox}, {Weaver},
  {Zabalza}, {Edwards}, {Azalee Bostroem}, {Burke}, {Casey}, {Crawford},
  {Dencheva}, {Ely}, {Jenness}, {Labrie}, {Lim}, {Pierfederici}, {Pontzen},
  {Ptak}, {Refsdal}, {Servillat}, \& {Streicher}}]{astropy2013}
{Astropy Collaboration}, {Robitaille}, T.~P., {Tollerud}, E.~J., {et~al.} 2013,
  \aap, 558, A33

\bibitem[{{Azimlu} {et~al.}(2011){Azimlu}, {Marciniak}, \&
  {Barmby}}]{azimlu2011}
{Azimlu}, M., {Marciniak}, R., \& {Barmby}, P. 2011, \aj, 142, 139

\bibitem[{{Barmby} {et~al.}(2006){Barmby}, {Ashby}, {Bianchi}, {Engelbracht},
  {Gehrz}, {Gordon}, {Hinz}, {Huchra}, {Humphreys}, {Pahre},
  {P{\'e}rez-Gonz{\'a}lez}, {Polomski}, {Rieke}, {Thilker}, {Willner}, \&
  {Woodward}}]{barmby2006}
{Barmby}, P., {Ashby}, M.~L.~N., {Bianchi}, L., {et~al.} 2006, \apjl, 650, L45

\bibitem[{{Breen} {et~al.}(2010){Breen}, {Caswell}, {Ellingsen}, \&
  {Phillips}}]{breen2010}
{Breen}, S.~L., {Caswell}, J.~L., {Ellingsen}, S.~P., \& {Phillips}, C.~J.
  2010, \mnras, 406, 1487

\bibitem[{{Brunthaler} {et~al.}(2005){Brunthaler}, {Reid}, {Falcke},
  {Greenhill}, \& {Henkel}}]{brunthaler2005}
{Brunthaler}, A., {Reid}, M.~J., {Falcke}, H., {Greenhill}, L.~J., \& {Henkel},
  C. 2005, Science, 307, 1440

\bibitem[{{Brunthaler} {et~al.}(2007){Brunthaler}, {Reid}, {Falcke}, {Henkel},
  \& {Menten}}]{brunthaler2007}
{Brunthaler}, A., {Reid}, M.~J., {Falcke}, H., {Henkel}, C., \& {Menten}, K.~M.
  2007, \aap, 462, 101

\bibitem[{{Calzetti} {et~al.}(2005){Calzetti}, {Kennicutt}, {Bianchi},
  {Thilker}, {Dale}, {Engelbracht}, {Leitherer}, {Meyer}, {Sosey}, {Mutchler},
  {Regan}, {Thornley}, {Armus}, {Bendo}, {Boissier}, {Boselli}, {Draine},
  {Gordon}, {Helou}, {Hollenbach}, {Kewley}, {Madore}, {Martin}, {Murphy},
  {Rieke}, {Rieke}, {Roussel}, {Sheth}, {Smith}, {Walter}, {White}, {Yi},
  {Scoville}, {Polletta}, \& {Lindler}}]{calzetti2005}
{Calzetti}, D., {Kennicutt}, Jr., R.~C., {Bianchi}, L., {et~al.} 2005, \apj,
  633, 871

\bibitem[{{Castangia} {et~al.}(2008){Castangia}, {Tarchi}, {Henkel}, \&
  {Menten}}]{castangia2008}
{Castangia}, P., {Tarchi}, A., {Henkel}, C., \& {Menten}, K.~M. 2008, \aap,
  479, 111

\bibitem[{{Churchwell} {et~al.}(1990){Churchwell}, {Walmsley}, \&
  {Cesaroni}}]{churchwell1990}
{Churchwell}, E., {Walmsley}, C.~M., \& {Cesaroni}, R. 1990, \aaps, 83, 119

\bibitem[{{Codella} {et~al.}(1995){Codella}, {Palumbo}, {Pareschi}, {Scappini},
  {Caselli}, \& {Attolini}}]{codella1995}
{Codella}, C., {Palumbo}, G.~G.~C., {Pareschi}, G., {et~al.} 1995, \mnras, 276,
  57

\bibitem[{{Cyganowski} {et~al.}(2013){Cyganowski}, {Koda}, {Rosolowsky},
  {Towers}, {Donovan Meyer}, {Egusa}, {Momose}, \&
  {Robitaille}}]{cyganowski2013}
{Cyganowski}, C.~J., {Koda}, J., {Rosolowsky}, E., {et~al.} 2013, \apj, 764, 61

\bibitem[{{Cyganowski} {et~al.}(2008){Cyganowski}, {Whitney}, {Holden},
  {Braden}, {Brogan}, {Churchwell}, {Indebetouw}, {Watson}, {Babler},
  {Benjamin}, {Gomez}, {Meade}, {Povich}, {Robitaille}, \&
  {Watson}}]{cyganowski2008}
{Cyganowski}, C.~J., {Whitney}, B.~A., {Holden}, E., {et~al.} 2008, \aj, 136,
  2391

\bibitem[{{Dale} \& {Helou}(2002)}]{dale2002}
{Dale}, D.~A., \& {Helou}, G. 2002, \apj, 576, 159

\bibitem[{{Darling}(2011)}]{darling2011}
{Darling}, J. 2011, \apjl, 732, L2

\bibitem[{{Darling}(2013)}]{darling2013}
---. 2013, \apjl, 777, L21

\bibitem[{{Darling} {et~al.}(2008){Darling}, {Brogan}, \&
  {Johnson}}]{darling2008}
{Darling}, J., {Brogan}, C., \& {Johnson}, K. 2008, \apjl, 685, L39

\bibitem[{{Darling} {et~al.}(2016){Darling}, {Gerard}, {Amiri}, \&
  {Lawrence}}]{darling2016}
{Darling}, J., {Gerard}, B., {Amiri}, N., \& {Lawrence}, K. 2016, \apj,
  submitted

\bibitem[Deguchi \& Watson(1989)]{deguchi1989} Deguchi, S., \& Watson, W. D.  1989, \apj, 340 L17

\bibitem[Elitzur(1992)]{elitzur1992} Elitzur, M.  1992, \araa, 30, 75 

\bibitem[Elitzur et al.(1991)]{elitzur1991} Elitzur, M., McKee, C. F., \& Hollenbach, D. J.   1991, \apj, 367, 333

\bibitem[{{Feigelson} \& {Babu}(2013)}]{feigelson2013}
{Feigelson}, E.~D., \& {Babu}, G.~J. 2013, {Statistical Methods for Astronomy},
  ed. T.~D. {Oswalt} \& H.~E. {Bond}, 445

\bibitem[{{Felli} {et~al.}(1992){Felli}, {Palagi}, \& {Tofani}}]{felli1992}
{Felli}, M., {Palagi}, F., \& {Tofani}, G. 1992, \aap, 255, 293

\bibitem[{{Ford} {et~al.}(2013){Ford}, {Gear}, {Smith}, {Eales}, {Baes},
  {Bendo}, {Boquien}, {Boselli}, {Cooray}, {De Looze}, {Fritz}, {Gentile},
  {Gomez}, {Gordon}, {Kirk}, {Lebouteiller}, {O'Halloran}, {Spinoglio},
  {Verstappen}, \& {Wilson}}]{ford2013}
{Ford}, G.~P., {Gear}, W.~K., {Smith}, M.~W.~L., {et~al.} 2013, \apj, 769, 55

\bibitem[{{Goddi} {et~al.}(2011){Goddi}, {Moscadelli}, \& {Sanna}}]{goddi2011}
{Goddi}, C., {Moscadelli}, L., \& {Sanna}, A. 2011, \aap, 535, L8

\bibitem[{{Goddi} {et~al.}(2006){Goddi}, {Moscadelli}, {Torrelles}, {Uscanga},
  \& {Cesaroni}}]{goddi2006}
{Goddi}, C., {Moscadelli}, L., {Torrelles}, J.~M., {Uscanga}, L., \&
  {Cesaroni}, R. 2006, \aap, 447, L9

\bibitem[{{Gordon} {et~al.}(2005){Gordon}, {Rieke}, {Engelbracht}, {Muzerolle},
  {Stansberry}, {Misselt}, {Morrison}, {Cadien}, {Young}, {Dole}, {Kelly},
  {Alonso-Herrero}, {Egami}, {Su}, {Papovich}, {Smith}, {Hines}, {Rieke},
  {Blaylock}, {P{\'e}rez-Gonz{\'a}lez}, {Le Floc'h}, {Hinz}, {Latter},
  {Hesselroth}, {Frayer}, {Noriega-Crespo}, {Masci}, {Padgett}, {Smylie}, \&
  {Haegel}}]{gordon2005}
{Gordon}, K.~D., {Rieke}, G.~H., {Engelbracht}, C.~W., {et~al.} 2005, \pasp,
  117, 503

\bibitem[{{Gordon} {et~al.}(2006){Gordon}, {Bailin}, {Engelbracht}, {Rieke},
  {Misselt}, {Latter}, {Young}, {Ashby}, {Barmby}, {Gibson}, {Hines}, {Hinz},
  {Krause}, {Levine}, {Marleau}, {Noriega-Crespo}, {Stolovy}, {Thilker}, \&
  {Werner}}]{gordon2006}
{Gordon}, K.~D., {Bailin}, J., {Engelbracht}, C.~W., {et~al.} 2006, \apjl, 638,
  L87

\bibitem[{{Greenhill} {et~al.}(1995){Greenhill}, {Henkel},
  {Becker}, {Wilson}, \& {Wouterloot}}]{greenhill1995}
{Greenhill}, L.~J., {Henkel}, C., {Becker}, R., {Wilson}, T.~L., \&
  {Wouterloot}, J.~G.~A. 1995, \aap, 304, 21

\bibitem[{{Groves} {et~al.}(2012){Groves}, {Krause}, {Sandstrom}, {Schmiedeke},
  {Leroy}, {Linz}, {Kapala}, {Rix}, {Schinnerer}, {Tabatabaei}, {Walter}, \&
  {da Cunha}}]{groves2012}
{Groves}, B., {Krause}, O., {Sandstrom}, K., {et~al.} 2012, \mnras, 426, 892

\bibitem[{{Henkel} {et~al.}(2005){Henkel}, {Peck}, {Tarchi}, {Nagar}, {Braatz},
  {Castangia}, \& {Moscadelli}}]{henkel2005}
{Henkel}, C., {Peck}, A.~B., {Tarchi}, A., {et~al.} 2005, \aap, 436, 75

\bibitem[{{Hollenbach} {et~al.}(2013){Hollenbach}, {Elitzur}, \&
  {McKee}}]{hollenbach2013}
{Hollenbach}, D., {Elitzur}, M., \& {McKee}, C.~F. 2013, \apj, 773, 70

\bibitem[{{Honma} {et~al.}(2005){Honma}, {Bushimata}, {Choi}, {Fujii},
  {Hirota}, {Horiai}, {Imai}, {Inomata}, {Ishitsuka}, {Iwadate}, {Jike},
  {Kameya}, {Kamohara}, {Kan-Ya}, {Kawaguchi}, {Kijima}, {Kobayashi}, {Kuji},
  {Kurayama}, {Manabe}, {Miyaji}, {Nakagawa}, {Nakashima}, {Oh}, {Omodaka},
  {Oyama}, {Rioja}, {Sakai}, {Sato}, {Sasao}, {Shibata}, {Shimizu}, {Sora},
  {Suda}, {Tamura}, \& {Yamashita}}]{honma2005}
{Honma}, M., {Bushimata}, T., {Choi}, Y.~K., {et~al.} 2005, \pasj, 57, 595

\bibitem[Humphreys et al.(2013)]{humphreys2013} Humphreys, E. M. L., 
Reid, M. J., Moran, J. M., Greenhill, L. J., \& Argon, A. L.   2013, \apj, 775, 13

\bibitem[{{Imai} {et~al.}(2001){Imai}, {Ishihara}, {Kameya}, \&
  {Nakai}}]{imai2001}
{Imai}, H., {Ishihara}, Y., {Kameya}, O., \& {Nakai}, N. 2001, \pasj, 53, 489

\bibitem[{{Jaffe} {et~al.}(1981){Jaffe}, {Guesten}, \& {Downes}}]{jaffe1981}
{Jaffe}, D.~T., {Guesten}, R., \& {Downes}, D. 1981, \apj, 250, 621

\bibitem[{{Kurtz} \& {Hofner}(2005)}]{kurtz2005}
{Kurtz}, S., \& {Hofner}, P. 2005, \aj, 130, 711

\bibitem[{{Laher} {et~al.}(2012){Laher}, {Gorjian}, {Rebull}, {Masci},
  {Fowler}, {Helou}, {Kulkarni}, \& {Law}}]{laher2012}
{Laher}, R.~R., {Gorjian}, V., {Rebull}, L.~M., {et~al.} 2012, \pasp, 124, 737

\bibitem[Liljestrom et al.(1989)]{liljestrom1989}Liljestrom, T., Mattila, K., 
Toriseva, M., \& Anttila, R.  1989, \aaps, 79, 19

\bibitem[Lo(2005)]{lo2005} Lo, K. Y.   2005, \araa, 43, 625 

\bibitem[Loeb et al.(2005)]{loeb2005} Loeb, A., Reid, M. J., Brunthaler, A., \& Falcke, H.  2005, \apj, 633, 894

\bibitem[{{Massey} {et~al.}(2006){Massey}, {Olsen}, {Hodge}, {Strong},
  {Jacoby}, {Schlingman}, \& {Smith}}]{massey2006}
{Massey}, P., {Olsen}, K.~A.~G., {Hodge}, P.~W., {et~al.} 2006, \aj, 131, 2478

\bibitem[{{Merrett} {et~al.}(2006){Merrett}, {Merrifield}, {Douglas},
  {Kuijken}, {Romanowsky}, {Napolitano}, {Arnaboldi}, {Capaccioli}, {Freeman},
  {Gerhard}, {Coccato}, {Carter}, {Evans}, {Wilkinson}, {Halliday}, \&
  {Bridges}}]{merrett2006}
{Merrett}, H.~R., {Merrifield}, M.~R., {Douglas}, N.~G., {et~al.} 2006, \mnras,
  369, 120

\bibitem[{{Molinari} {et~al.}(2008){Molinari}, {Pezzuto}, {Cesaroni}, {Brand},
  {Faustini}, \& {Testi}}]{molinari2008}
{Molinari}, S., {Pezzuto}, S., {Cesaroni}, R., {et~al.} 2008, \aap, 481, 345

\bibitem[{{Moscadelli} {et~al.}(2011){Moscadelli}, {Cesaroni}, {Rioja},
  {Dodson}, \& {Reid}}]{moscadelli2011}
{Moscadelli}, L., {Cesaroni}, R., {Rioja}, M.~J., {Dodson}, R., \& {Reid},
  M.~J. 2011, \aap, 526, A66

\bibitem[{{Moscadelli} {et~al.}(2006){Moscadelli}, {Testi}, {Furuya}, {Goddi},
  {Claussen}, {Kitamura}, \& {Wootten}}]{moscadelli2006}
{Moscadelli}, L., {Testi}, L., {Furuya}, R.~S., {et~al.} 2006, \aap, 446, 985

\bibitem[{{Mould} {et~al.}(2008){Mould}, {Barmby}, {Gordon}, {Willner},
  {Ashby}, {Gehrz}, {Humphreys}, \& {Woodward}}]{mould2008}
{Mould}, J., {Barmby}, P., {Gordon}, K., {et~al.} 2008, \apj, 687, 230

\bibitem[{{Nieten} {et~al.}(2006){Nieten}, {Neininger}, {Gu{\'e}lin},
  {Ungerechts}, {Lucas}, {Berkhuijsen}, {Beck}, \& {Wielebinski}}]{nieten2006}
{Nieten}, C., {Neininger}, N., {Gu{\'e}lin}, M., {et~al.} 2006, \aap, 453, 459

\bibitem[{{Palla} {et~al.}(1991){Palla}, {Brand}, {Comoretto}, {Felli}, \&
  {Cesaroni}}]{palla1991}
{Palla}, F., {Brand}, J., {Comoretto}, G., {Felli}, M., \& {Cesaroni}, R. 1991,
  \aap, 246, 249

\bibitem[{{Piazzo}(2013)}]{piazzo2013}
{Piazzo}, L. 2013, ArXiv e-prints, arXiv:1301.1246

\bibitem[{{Pilbratt} {et~al.}(2010){Pilbratt}, {Riedinger}, {Passvogel},
  {Crone}, {Doyle}, {Gageur}, {Heras}, {Jewell}, {Metcalfe}, {Ott}, \&
  {Schmidt}}]{pilbratt2010}
{Pilbratt}, G.~L., {Riedinger}, J.~R., {Passvogel}, T., {et~al.} 2010, \aap,
  518, L1

\bibitem[{{Poglitsch} {et~al.}(2010){Poglitsch}, {Waelkens}, {Geis},
  {Feuchtgruber}, {Vandenbussche}, {Rodriguez}, {Krause}, {Renotte}, {van
  Hoof}, {Saraceno}, {Cepa}, {Kerschbaum}, {Agn{\`e}se}, {Ali}, {Altieri},
  {Andreani}, {Augueres}, {Balog}, {Barl}, {Bauer}, {Belbachir}, {Benedettini},
  {Billot}, {Boulade}, {Bischof}, {Blommaert}, {Callut}, {Cara}, {Cerulli},
  {Cesarsky}, {Contursi}, {Creten}, {De Meester}, {Doublier}, {Doumayrou},
  {Duband}, {Exter}, {Genzel}, {Gillis}, {Gr{\"o}zinger}, {Henning},
  {Herreros}, {Huygen}, {Inguscio}, {Jakob}, {Jamar}, {Jean}, {de Jong},
  {Katterloher}, {Kiss}, {Klaas}, {Lemke}, {Lutz}, {Madden}, {Marquet},
  {Martignac}, {Mazy}, {Merken}, {Montfort}, {Morbidelli}, {M{\"u}ller},
  {Nielbock}, {Okumura}, {Orfei}, {Ottensamer}, {Pezzuto}, {Popesso},
  {Putzeys}, {Regibo}, {Reveret}, {Royer}, {Sauvage}, {Schreiber}, {Stegmaier},
  {Schmitt}, {Schubert}, {Sturm}, {Thiel}, {Tofani}, {Vavrek}, {Wetzstein},
  {Wieprecht}, \& {Wiezorrek}}]{poglitsch2010}
{Poglitsch}, A., {Waelkens}, C., {Geis}, N., {et~al.} 2010, \aap, 518, L2

\bibitem[Reid \& Moran(1981)]{reid1981} Reid, M. J. \& Moran, J. M.  1981, \araa, 19, 231 

\bibitem[Reid \& Honma(2014)]{reid2014} Reid, M. J. \& Honma, M.  2014, \araa, 52, 339

\bibitem[{{S{\'a}nchez-Monge} {et~al.}(2013){S{\'a}nchez-Monge}, {Beltr{\'a}n},
  {Cesaroni}, {Fontani}, {Brand}, {Molinari}, {Testi}, \&
  {Burton}}]{sanchez2013}
{S{\'a}nchez-Monge}, {\'A}., {Beltr{\'a}n}, M.~T., {Cesaroni}, R., {et~al.}
  2013, \aap, 550, A21

\bibitem[{{Seth} {et~al.}(2002){Seth}, {Greenhill}, \& {Holder}}]{seth2002}
{Seth}, A.~C., {Greenhill}, L.~J., \& {Holder}, B.~P. 2002, \apj, 581, 325

\bibitem[{{Smith} {et~al.}(2012){Smith}, {Eales}, {Gomez}, {Roman-Duval},
  {Fritz}, {Braun}, {Baes}, {Bendo}, {Blommaert}, {Boquien}, {Boselli},
  {Clements}, {Cooray}, {Cortese}, {de Looze}, {Ford}, {Gear}, {Gentile},
  {Gordon}, {Kirk}, {Lebouteiller}, {Madden}, {Mentuch}, {O'Halloran}, {Page},
  {Schulz}, {Spinoglio}, {Verstappen}, {Wilson}, \& {Thilker}}]{smith2012}
{Smith}, M.~W.~L., {Eales}, S.~A., {Gomez}, H.~L., {et~al.} 2012, \apj, 756, 40

\bibitem[{{Sohn} {et~al.}(2012){Sohn}, {Anderson}, \& {van der
  Marel}}]{sohn2012}
{Sohn}, S.~T., {Anderson}, J., \& {van der Marel}, R.~P. 2012, \apj, 753, 7

\bibitem[{{Sullivan}(1973)}]{sullivan1973}
{Sullivan}, III, W.~T. 1973, \apjs, 25, 393

\bibitem[{{Urquhart} {et~al.}(2011){Urquhart}, {Morgan}, {Figura}, {Moore},
  {Lumsden}, {Hoare}, {Oudmaijer}, {Mottram}, {Davies}, \&
  {Dunham}}]{urquhart2011}
{Urquhart}, J.~S., {Morgan}, L.~K., {Figura}, C.~C., {et~al.} 2011, \mnras,
  418, 1689

\bibitem[{{van der Marel} {et~al.}(2012){van der Marel}, {Fardal}, {Besla},
  {Beaton}, {Sohn}, {Anderson}, {Brown}, \& {Guhathakurta}}]{vandermarel2012}
{van der Marel}, R.~P., {Fardal}, M., {Besla}, G., {et~al.} 2012, \apj, 753, 8

\bibitem[{{Walker} {et~al.}(1982){Walker}, {Matsakis}, \&
  {Garcia-Barreto}}]{walker1982}
{Walker}, R.~C., {Matsakis}, D.~N., \& {Garcia-Barreto}, J.~A. 1982, \apj, 255,
  128

\bibitem[{{Walsh} {et~al.}(2011){Walsh}, {Breen}, {Britton}, {Brooks},
  {Burton}, {Cunningham}, {Green}, {Harvey-Smith}, {Hindson}, {Hoare},
  {Indermuehle}, {Jones}, {Lo}, {Longmore}, {Lowe}, {Phillips}, {Purcell},
  {Thompson}, {Urquhart}, {Voronkov}, {White}, \& {Whiting}}]{walsh2011}
{Walsh}, A.~J., {Breen}, S.~L., {Britton}, T., {et~al.} 2011, \mnras, 416, 1764

\bibitem[{{Wilson}(1927)}]{wilson1927}
{Wilson}, E. B.  1927, Journal of the American Statistical Association, 22, 158, 209

\bibitem[{{Wright} {et~al.}(2010){Wright}, {Eisenhardt}, {Mainzer}, {Ressler},
  {Cutri}, {Jarrett}, {Kirkpatrick}, {Padgett}, {McMillan}, {Skrutskie},
  {Stanford}, {Cohen}, {Walker}, {Mather}, {Leisawitz}, {Gautier}, {McLean},
  {Benford}, {Lonsdale}, {Blain}, {Mendez}, {Irace}, {Duval}, {Liu}, {Royer},
  {Heinrichsen}, {Howard}, {Shannon}, {Kendall}, {Walsh}, {Larsen}, {Cardon},
  {Schick}, {Schwalm}, {Abid}, {Fabinsky}, {Naes}, \& {Tsai}}]{wright2010}
{Wright}, E.~L., {Eisenhardt}, P.~R.~M., {Mainzer}, A.~K., {et~al.} 2010, \aj,
  140, 1868

\end{thebibliography}
\end{document}